\documentclass[%
 reprint,
nofootinbib,
  showpacs,
 amsmath,amssymb,
 aps,
prb,
longbibliography,
superscriptaddress,
]{revtex4-2}

\usepackage[dvipdfmx]{graphicx}
\usepackage{bmpsize} 
\usepackage{dcolumn}
\usepackage{pifont} 
\usepackage{bm}
\usepackage{multirow}
\usepackage{longtable}
\usepackage[utf8]{inputenc}   
\usepackage[T1]{fontenc}      
\usepackage{lmodern}          
\usepackage{csquotes}         
\usepackage{amsmath}          
\usepackage{mathtools}        
\usepackage{amstext}         
\usepackage{amssymb}          
\usepackage{subdepth}         
\usepackage{xcolor}         
\usepackage[
  separate-uncertainty=true,
  per-mode=reciprocal,
  binary-units=true,
  range-phrase=--]{siunitx}  
\usepackage{braket}           %
\usepackage{graphicx}         
\usepackage[pass]{geometry}  
\usepackage[capitalize]{cleveref}

\usepackage{times}
\usepackage{booktabs}

\begin{document}

\preprint{APS/123-QED}

\title{
Generation of modulated magnetic structure  
 based on cluster multipole: \\
 Application to $\alpha$-Mn and Co$M_3$S$_6$ 
}

\author{Yuki Yanagi}
\affiliation{Center for Computational Materials Science, Institute for Materials Research, Tohoku University, Sendai, Miyagi, 950-8577, Japan}
\author{Hiroaki Kusunose}
\affiliation{Department of Physics, Meiji University, Kawasaki 214-8571, Japan}
\author{Takuya Nomoto}
\affiliation{University of Tokyo, 7-3-1 Hongo, Bunkyo-ku, Tokyo 113-8656, Japan}
\author{Ryotaro Arita}
\affiliation{University of Tokyo, 7-3-1 Hongo, Bunkyo-ku, Tokyo 113-8656, Japan}
\affiliation{Center for Emergent Matter Science, RIKEN, Wako, Saitama 351-0198, Japan}
\author{Michi-To Suzuki}
\affiliation{Center for Computational Materials Science, Institute for Materials Research, Tohoku University, Sendai, Miyagi, 950-8577, Japan}
\affiliation{Center for Spintronics Research Network, Graduate School of Engineering Science, Osaka University, Toyonaka, Osaka 560-8531, Japan}

\begin{abstract}
  We present a systematic method to automatically generate symmetry-adapted magnetic structures for given crystal structure and  general propagation vector ${\bm k}$ as an efficient approach of the analysis of complex modulated magnetic structures.
 The method is developed as an extension of the generation scheme based on multipole expansion, which was demonstrated only for the propagation vector ${\bm k}=\bm{0}$ [M.-T. Suzuki \textit{et al.,} Phys. Rev. B \textbf{99}, 174407 (2019)]. 
  The symmetry-adapted magnetic structures characterized with an ordering vector ${\bm k}$ are obtained by mapping the multipole magnetic alignments on a virtual cluster to the periodic crystal structure with the phase factor for the wave vector ${\bm k}$.
   This method provides all magnetic bases compatible with irreducible representations under $\bm{k}$-group for given crystal structure and wave vector $\bm{k}$. 
    Multiple-$\bm{k}$ magnetic structures are derived from  
    superposition of 
    single-$\bm{k}$ magnetic bases related with space group symmetry. 
    We apply the scheme to deduce the magnetic structures of $\alpha$-Mn and  
    Co$M_3$S$_6$ ($M$~=~Nb, Ta), in which large anomalous Hall effect has recently been observed in antiferromagnetic phases,  
     and identify the magnetic structures inducing anomalous Hall effect without net magnetization. 
 The physical phenomena originating from emergent multipoles in the ordered phases are also discussed based on the Landau theory.
\end{abstract}

\maketitle


\section{Introduction}

Magnetic materials with complexly arranged magnetic moments in crystal  
  provide a platform for exploring various exotic phenomena and potential device applications. 
The details of magnetic structures play a crucial role for intriguing physical phenomena, such as anomalous/topological Hall effect, non-reciprocal charge transport, and magnetoelectric effect. 
 Therefore, the characterization of magnetic structures is a key to understanding such physical phenomena in magnetically ordered phases.

 A systematic generation and characterization scheme of symmetry-adapted basis set based on the multipole expansion combined with group theory has been proposed in our previous studies~\cite{Suzuki_PhysRevB.95.094406,Suzuki_PhysRevB.99.174407}. 
 In this method, multipole expansion for macroscopic electromagnetic fields is extended to describe the configuration of magnetic moments in general magnetic orderings using multipolar moments with IRREPs under crystallographic point group. 
  This enables us with an efficient analysis of electronic properties and macroscopic responses in antiferromagnets. It has been elucidated that the magnetic structure in a noncollinear antiferromagnet Mn$_3$Sn  can be viewed as a magnetic octupole with use of this method~\cite{Suzuki_PhysRevB.95.094406,Suzuki_JPSJ.87.041008}.
   The magnetic octupole scenario provides a unified understanding of anomalous responses in Mn$_3$Sn such as large anomalous Hall effect (AHE)~\cite{Nakatsuji_Mn3Sn_AHE_2015}, anomalous Nernst effect (ANE)~\cite{Ikhlas_Nat.Physics.13.1085}, 
    and magneto-optical Kerr effect (MOKE)~\cite{Higo_Nat.Photonics.12.73}. 
    The analysis based on multipoles deepens insights into AHE, ANE, and MOKE  beyond the conventional understanding that these phenomena are induced by net magnetization with spin-orbit coupling~\cite{Karplus_PhysRev.95.1154}.

   There are also advantages of using symmetry-classified magnetic structures in respect to prediction for  stable and/or meta-stable magnetic states.  
  Many experimental studies imply that most of magnetic structures are specified by a small number of irreducible representation (IRREP)~\cite{Gallego:ks5532}, which classify possible transformation property of magnetic structures for the space group operation of crystal~\cite{Bertaut:a05871,Izyumov_magnetic.rep,Bradley_Cracknell_1972}. 
 Actually, the recent benchmark calculation using high-throughput prediction scheme of magnetic structure making use of  multipole expansion shows that the symmetrically classified magnetic structures are efficient candidate of initial magnetic structures employed in the first-principles   calculations for given crystal structures~\cite{Huebsch_PhysRevX.11.011031}. 

In our magnetic structure generation method proposed previously~\cite{Suzuki_PhysRevB.95.094406,Suzuki_PhysRevB.99.174407,Huebsch_PhysRevX.11.011031}, we only considered magnetic structures 
which have no spatial modulation over inter-unit cell, that is, magnetic structures having propagation vector $\bm{k}=\bm{0}$. 
Meanwhile, many of attracting magnetic structures  such as helical magnetic order, spin density wave, and skyrmion crystals have finite propagation vector ${\bm k}$~\cite{Nagaosa_Nat.Nanotech.8.899}.
In this paper, we develop  a generation scheme of symmetry-adapted magnetic structures 
 with inter-unit cell spatial modulation quantified by  propagation vector ${\bm k}$, along the line of the scheme proposed in Refs.~\cite{Suzuki_PhysRevB.95.094406,Suzuki_PhysRevB.99.174407}. 
 Our algorithm based on multipole expansion largely reduces the degree of freedom of generated magnetic structures and  efficiently provides magnetic structures symmetrically adapted to the crystal structures with getting around under- and over-generation of basis set, that produces too few or too many basis vectors~\cite{Stokes_PhysRevB.43.11010,Davies_Hindawi}. 
  The method  paves a way to explore the intricate magnetic structure  by making full use of symmetry information. 

In Sec.~II, we provide general formulation of the generation scheme of symmetry adapted magnetic structures with finite ${\bm k}$ vectors. We introduce multipole expansion of the magnetic structure on an atomic cluster in Sec.~II A and discuss how to define the virtual atomic cluster on which the multipole expansion is defined in Sec.~II B. We then show how to generate symmetry adapted magnetic structures with single wave vector (Sec.~II C) and with multiple wave vectors (Sec.~II D).  
 We further discuss possible magnetic structures of $\alpha$-Mn and Co$M_3$S$_6$  ($M$~=~Nb, Ta), which are known to exhibit large anomalous Hall responses in the AFM phases, by applying the generation scheme of magnetic structures in Sec.~III. 
 Finally, we summarize the paper and give a future perspective for design of magnetic materials in Sec.~IV.

\section{Methods}

\subsection{Multipole expansion}

 We first review the multipole expansion of vector potential $\bm{A}(\bm{r})$ and its application to magnetic structure introduced in Ref.~\cite{Suzuki_PhysRevB.99.174407}. 
   As is well known,  a spatial distribution of vector potential $\bm{A}(\bm{r})$ outside its sources is 
    systematically characterized by magnetic (M) and magnetic toroidal (MT) multipole moments, $M_{l m}$ and $T_{l m}$, as follows~\cite{Jackson_classical.elect.dynam,Schwartz_PhysRev.97.380,DUBOVIK1990145}: 
\begin{align}
\bm{A}\left(\bm{r}\right) 
&= \sum_{lm}
\Biggl[\sqrt{\frac{4\pi\left(l + 1 \right)}{\left(2l + 1\right)l}} M_{l m} \frac{\bm{Y}^{l}_{l m}\left(\hat{\bm{r}}\right)}{ir^{l + 1}}  \nonumber \\
&-\sqrt{ 4\pi\left(l + 1 \right)} T_{l m} \frac{\bm{Y}^{l+1}_{l m}\left(\hat{\bm{r}}\right)}{r^{l + 2}}
\Biggr],  \label{eq_mp_exp}
\end{align}
where the Coulomb gauge $\bm{\nabla}\cdot \bm{A}(\bm{r})=0$ is imposed and $\bm{Y}^{l'}_{lm}(\hat{\bm{r}})$ ($l\ge 1$, $-l\le m \le l$, $l'=l-1,l,l+1$) represents the vector spherical harmonics with  $\hat{\bm{r}}=\bm{r}/r$~\cite{Varshalovich_vector.spher,Kusunose_JPSJ.77.064710}. 
The M and MT multipole moments are given as,
\begin{align}
&M_{lm} = \sum_{j} \left( \frac{2\bm{l}_j}{l+1} + \bm{\sigma}_j\right) \cdot  \bm{O}_{lm}(\bm{r}_j), \label{eq_M_mp_O(3)} \\ 
&T_{lm} =  \sum_{j} \left\{\frac{\bm{r}_j}{l+1} \times \left( \frac{2\bm{l}_j}{l+2} + \bm{\sigma}_j\right) \right\} \cdot  \bm{O}_{lm}(\bm{r}_j), \label{eq_T_mp_O(3)}
\end{align}
with 
\begin{align}
  \bm{O}_{lm}(\bm{r}) = \sqrt{\frac{4\pi}{2l+1} } \bm{\nabla} [r^l Y^*_{lm} (\hat{\bm{r}})], \label{eq_Olm}
\end{align}
where $\bm{l}_j$ ($\bm{\sigma}_j$) and $\bm{r}_j$ are orbital (spin) angular momentum and position vector of each electron, respectively, and $ \bm{\nabla}(r^{l}Y_{lm}) =  r^{l-1}\sqrt{l(2l+1)} \bm{Y}^{l-1}_{lm} $ 
 with parity for spatial inversion  $(-1)^{l+1}$. 
 We define M- and MT-multipole moments in an atomic cluster by replacing the electron coordinate $\bm{r}_j$ by  atomic coordinate  of $j$-th atom $\bm{\xi}_{j}$ and neglecting the orbital magnetic moment in Eqs.~(\ref{eq_M_mp_O(3)}) and (\ref{eq_T_mp_O(3)}) 
  as discussed in Ref.~\cite{Suzuki_PhysRevB.99.174407}.  
 Here and hereafter, the electric- (E-),  M-, electric toroidal- (ET-), and 
  MT-multipoles are denoted by $Q$, $M$, $G$, and $T$, respectively 
   according to the notations in Ref.~\cite{Hayami_PhysRevB.98.165110}.
 Taking  into consideration that the atomic cluster is not invariant under continuous rotation but have a point group symmetry, 
the magnetic bases $\bm{\psi}^{(X)}_{l\Gamma\gamma }$ on the cluster characterized by symmetry-adapted multipole moments  are  obtained as follows: 
 \begin{align}
\bm{\psi}^{(X)}_{l\Gamma\gamma } 
&=\sum^{N}_{j=1}\sum_{\mu=x,y,z} u^{(X)}_{l\Gamma\gamma,j\mu} \bm{e}^{\mathrm{axial}}_{j\mu} \nonumber \\
&= \sum^{N}_{j=1} \bm{u}^{(X)}_{l\Gamma\gamma,j}\cdot \bm{e}^{\mathrm{axial}}_{j},  ~\quad  (X=M,T),  \label{eq_mag_clust}
 \end{align}
   where  $\bm{e}^{\mathrm{axial}}_{j\mu}$ denotes the axial unit vector located on $j$-th atom along $\mu(=x,y,z)$-direction with odd parity under time-reversal operation. 
   Each magnetic basis $\bm{\psi}^{(X)}_{l\Gamma\gamma }$ is defined as $3N$-dimensional vector. 
 The coefficients $u^{(X)}_{l\Gamma\gamma,j\mu}$ are given as follow:  
\begin{align}
 & \bm{u}^{(M)}_{l\Gamma\gamma,j} =   \bm{\mathcal{O}}_{l\Gamma\gamma}(\bm{\xi}_j), \label{eq_M_mp_pg} \\
 & \bm{u}^{(T)}_{l\Gamma\gamma,j} =  \frac{1}{l+1}  \bm{\mathcal{O}}_{l\Gamma\gamma}(\bm{\xi}_j) \times \bm{\xi}_j,\label{eq_T_mp_pg}  
 \end{align}
 where   $\Gamma$ and $\gamma$ represent IRREPs of point group and its component, respectively and 
 $\bm{\mathcal{O}}_{l\Gamma\gamma}$ is obtained by replacing $Y_{lm}$ in   Eq.~(\ref{eq_Olm})
  by symmetry-adapted function with $\Gamma$-IRREP  in the point group of a given atomic cluster, which can be represented as linear combination of  $Y_{lm}$ with fixed rank $l$. 
  For instance, if we replace $Y_{2m}$ by $\mathcal{Y}_{\Gamma\gamma}\propto x^2-y^2 \propto Y_{2-2}+ Y_{22}$, we obtain M (MT)-multipole $\bm{u}^{(X)}_{l\Gamma\gamma,j}$ ($X=M,l=2,\Gamma=B_{1u}$)  [($X=T,l=2,\Gamma=B_{1g}$)]
  with one-dimensional $B_{1u}$ ($B_{1g}$)-IRREP under $D_{4h}$ point group.

\subsection{Virtual atomic cluster for multipole expansion}
\begin{figure}[t]
    \centering
    \includegraphics[width=1.0 \hsize]{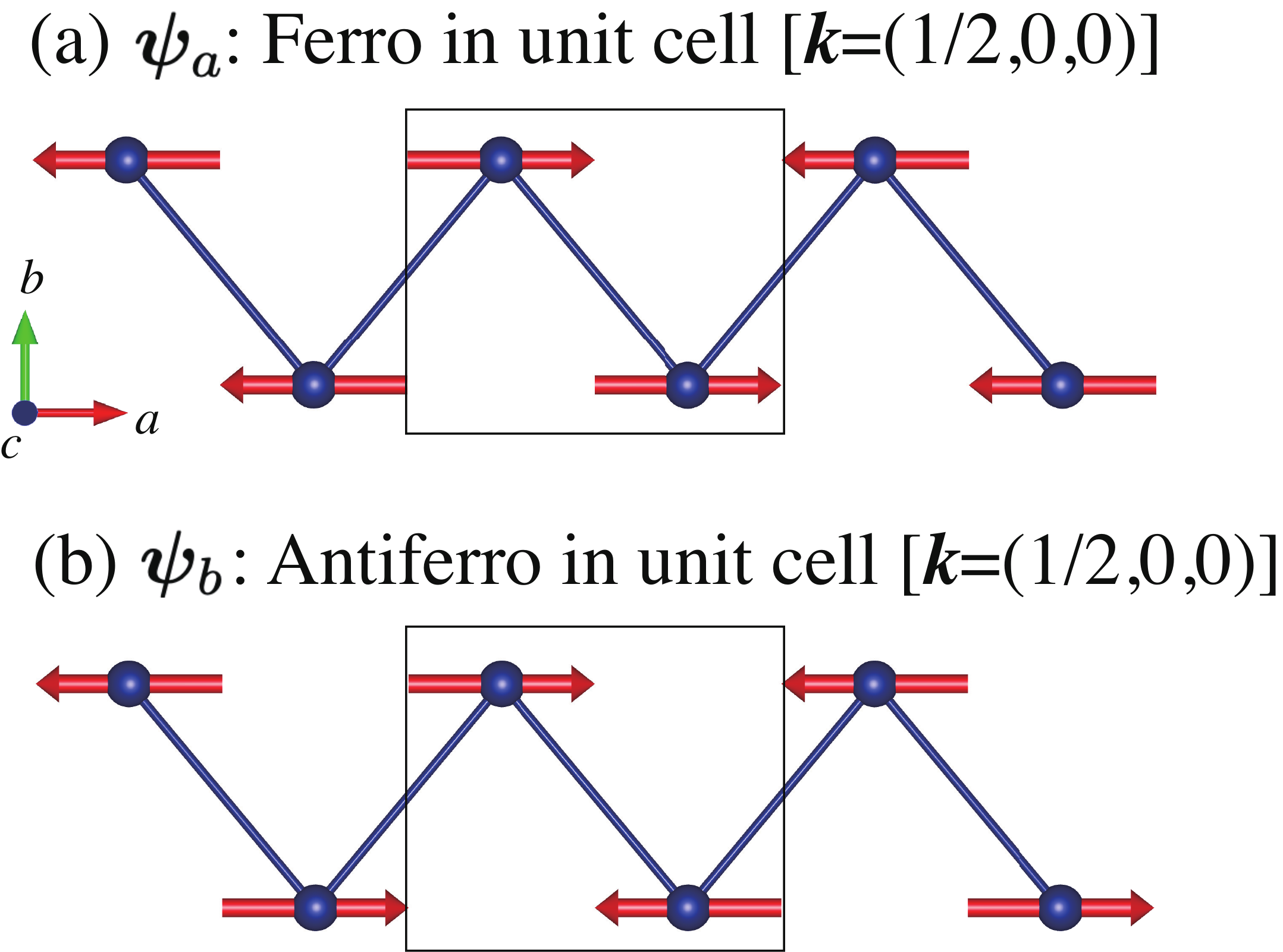}
    \caption{(a) Ferro and (b) antiferro magnetic structures in unit cell with $\bm{k}=(\frac{1}{2},0,0)$ on zigzag chain.}
    \label{fig:mag_zigzag}
\end{figure}

  We formulated schemes to generate symmetry-adapted orthogonal magnetic structures without spacial modulation of the alignment, i.e. $\bm{k}={\bm 0}$, in earlier work~\cite{Suzuki_PhysRevB.95.094406,Suzuki_PhysRevB.99.174407}.
  In the method, we first define a virtual cluster consisting of non-overlapped atoms related to each other by the rotation operations of the crystallographic point group.  The crystallographic point group $\mathcal{P}$  is defined as
\begin{align}
  \mathcal{P}=\sum_{i=1}^{N_{\mathrm{coset}}}  \left\{p_{i}|\bm{0}\right\}  \mathcal{H}, 
\end{align}
for the space group $\mathcal{G}$ given as follows:  
  \begin{align}
\mathcal{G}=\sum_{i=1}^{N_{\mathrm{coset}}}  \left\{p_{i}|\bm{\tau}_{i}\right\}  \mathcal{H}\mathcal{T}, 
\end{align}
where $p_i$ represents the point group operation with $p_1=E$ being identity operation, $\bm{\tau}_i$ denotes non-primitive translation with $\bm{\tau}_1=\bm{0}$ and $\bm{\tau}_i\neq \bm{0}$ ($i\ge 2$),   and $\mathcal{T}$ is the translation group. 
 The subgroup of the space group $\mathcal{H}$ is  composed only of the rotational operations 
 of the space group, i.e., 
$\{h_{\zeta} | \bm{0} \} \in \mathcal{H} \subset \mathcal{G}$,  with $h_1=E$  $(\zeta=1,2,\ldots,N_{h})$. 
   In Ref.~\cite{Suzuki_PhysRevB.99.174407}, a virtual cluster is composed of $N_0\equiv N_{\mathrm{coset}}N_h$ sites whose positions are defined by operating point group operations  $p_{i}h_{\zeta}$ ($i=1,2,\ldots,N_{\mathrm{coset}}$, $\zeta=1,2,\ldots,N_{h}$) on a position vector $\bm{r} = \bm{r}_1$ 
   to classify magnetic structures having $\bm{k}=\bm{0}$ according to the IRREPs of point group.  
   For the case of finite propagation vector $\bm{k}$, there are specific IRREPs which are not equivalent to IRREPs in the point group symmetry of $\bm{k}$-group at the Brillouin zone boundary, associated with sublattice degrees of freedom by non-symmorphic space group operations.  
   For instance, the magnetic structures on zigzag chain with $\bm{k}=(\frac{1}{2},0,0)$ shown in Figs.~\ref{fig:mag_zigzag}(a) and (b) are obviously equivalent each other. 
 Note that zigzag structure has symmetry operations shown in Table~\ref{tab:zigzag} and the $\bm{k}$-point group at $\bm{k}=(\frac{1}{2},0,0)$ is $D_{2h}$. The magnetic structures shown in Figs.~\ref{fig:mag_zigzag} (a) and (b) are transformed to each other by symmetry operations $\{C_{2y}|(\frac{1}{2},0,0)\}$, $\{C_{2x}|(\frac{1}{2},0,0)\}$,   $\{IC_{2y}|(\frac{1}{2},0,0)\}$, and $\{IC_{2x}|(\frac{1}{2},0,0)\}$ according to Table~\ref{tab:zigzag}. 
 The other independent magnetic bases, in which the magnetic moments are directed along $b$- and $c$-axes, have similar transformation properties. 
  Therefore, these magnetic structures form two-dimensional IRREPs inequivalent to the IRREPs of $D_{2h}$ point group, all of which are one-dimensional.  
      In this case, 
     we cannot directly apply the generation scheme using virtual cluster for $\bm{k}=\bm{0}$ constructed according to point group operations. 
     To address this issue, we define  $N_{\mathrm{coset}}$-virtual clusters of which each cluster consists of  $N_{h}$-atoms for the magnetic structure generation for finite propagation vector $\bm{k}$ as discussed below. 

\begin{table}[b]
    \centering
    \caption{Representative elements for space group in zigzag chain and transformation properties of magnetic structures shown in Figs.~\ref{fig:mag_zigzag}, where $\bm{\psi}_a$ and $\bm{\psi}_b$ denote magnetic states in Figs.~\ref{fig:mag_zigzag} (a) and (b), respectively. The nonprimitive translation $\bm{\tau}$ is given by $ \bm{\tau} = (\frac{1}{2},0,0)$. }
    \label{tab:zigzag}
    \begin{tabular}{ccccc}
    \hline
    \hline
     & $\{E|\bm{0}\}$ & $\{C_{2z}|\bm{0}\}$ & $\{C_{2y}|\bm{\tau}\}$ & $\{C_{2x}|\bm{\tau}\}$ \\ \hline
    $\bm{\psi}_a$   & $\bm{\psi}_a$ & $-\bm{\psi}_a$ & $\bm{\psi}_b$ & $-\bm{\psi}_b$ \\
    $\bm{\psi}_b$   & $\bm{\psi}_b$ & $\bm{\psi}_b$ & $\bm{\psi}_a$ & $\bm{\psi}_a$  \\ \hline
     & $\{I|\bm{0}\}$ &      $\{IC_{2z}|\bm{0}\}$ & $\{IC_{2y}|\bm{\tau}\}$ & $\{IC_{2x}|\bm{\tau}\}$  \\   \hline
    $\bm{\psi}_a$    & $\bm{\psi}_a$ & $-\bm{\psi}_a$ & $\bm{\psi}_b$ & $-\bm{\psi}_b$ \\
    $\bm{\psi}_b$   & $-\bm{\psi}_b$ & $-\bm{\psi}_b$ & $-\bm{\psi}_a$ & $-\bm{\psi}_a$ \\ 
    \hline
    \hline
    \end{tabular}
\end{table}

 \subsection{Magnetic structure generation for single-$\bm{k}$ states}

\begin{figure}
    \centering
    \includegraphics[width=1.0 \hsize]{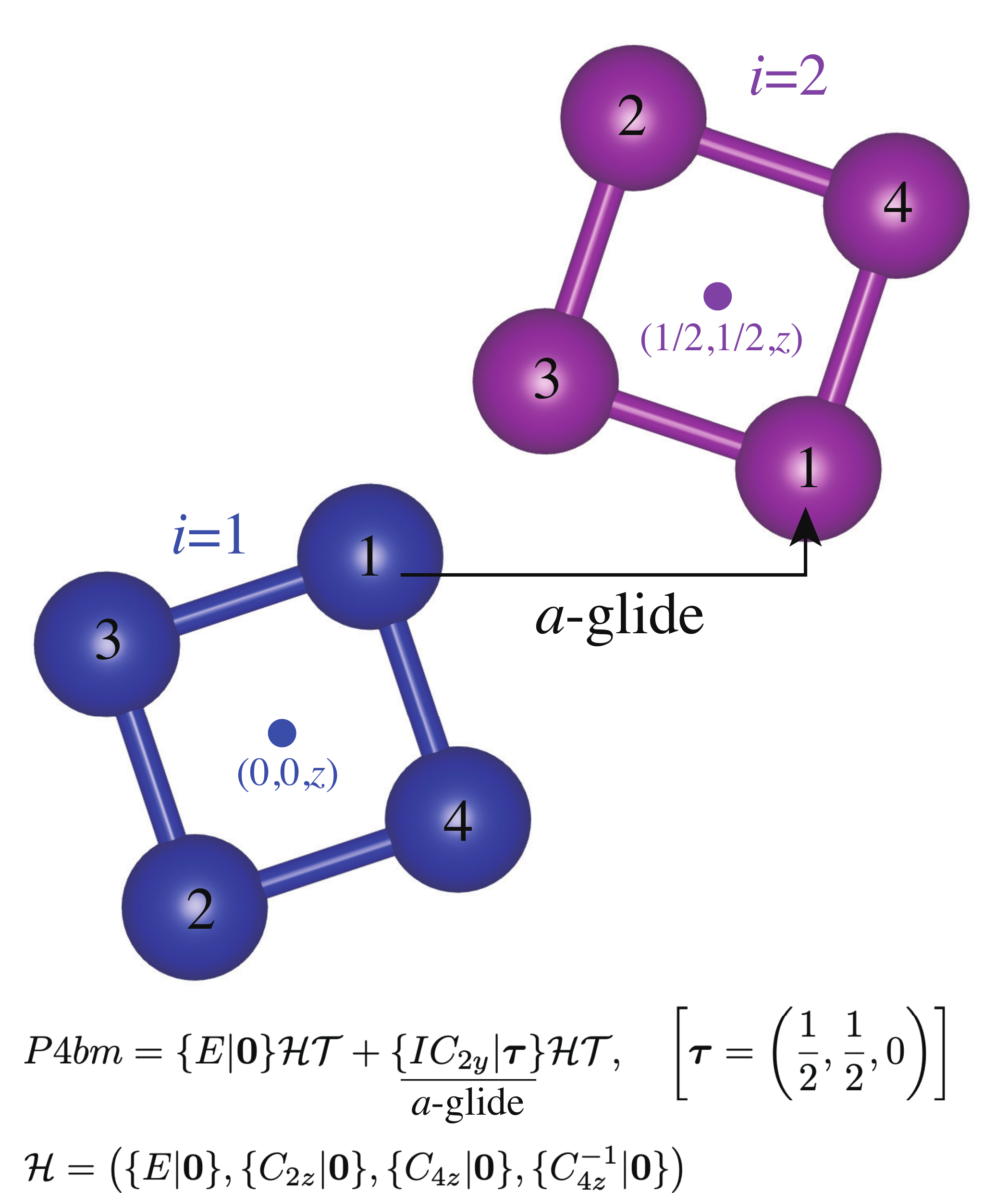}
    \caption{Virtual clusters in space group $P4bm$.}
    \label{fig:cluster}
\end{figure}

\begin{table}[t]
    \centering
    \caption{Atomic 
     indices 
    transformed by representative elements for space group $P4bm$. }
    \label{tab:cluster}
    \begin{tabular}{ccccc}
    \hline
    \hline
     & $\{E|\bm{0}\}$ & $\{C_{2z}|\bm{0}\}$ & $\{C_{4z}|\bm{0}\}$ & $\{C^{-1}_{4z}|\bm{0}\}$ \\ \hline
    $i$   & $1$ & $1$ & $1$ & $1$ \\
    $\zeta$   & $1$ & $2$ & $3$ & $4$  \\ 
    $s_{i_{\zeta}}[1]$   & $1$ & $1$ & $2$ & $2$  \\ 
    $s_{i_{\zeta}}[2]$   & $2$ & $2$ & $1$ & $1$  \\ \hline
     & $\{IC_{2y}|\bm{\tau}\}$ &      $\{IC_{2x}|\bm{\tau}\}$ & $\{IC_{2[1\bar{1}0]}|\bm{\tau}\}$ & $\{IC_{2[110]}|\bm{\tau}\}$  \\   \hline
    $i$    & $2$ & $2$ & $2$ & $2$ \\
    $\zeta$   & $1$ & $2$ & $3$ & $4$ \\ 
    $s_{i_{\zeta}}[1]$   & $2$ & $2$ & $1$ & $1$  \\ 
    $s_{i_{\zeta}}[2]$   & $1$ & $1$ & $2$ & $2$  \\ 
    \hline
    \hline
    \end{tabular}
\end{table}

 We first focus on single-$\bm{k}$ magnetic structure, whose periodicity is characterized by a single wave vector $\bm{k}$.  
  At a given wave vector $\bm{k}$, the magnetic bases are classified into the IRREPs of the little group of $\bm{k}$ denoted as  $\mathcal{G}_{\bm{k}}$. 
 The group elements in $\mathcal{G}_{\bm{k}}$   are composed of a subset of symmetry operations  in $\mathcal{G}$ which keeps $\bm{k}$ invariant under transformations. 
 $\mathcal{G}_{\bm{k}}$ is thus a subgroup of a space group $\mathcal{G}$. 
  We construct 
  the $N_{\mathrm{coset}}$-virtual clusters, where each of them have $N_{h}$-atoms. 
   By  operating group elements  $\left\{p_{i}|\bm{\tau}_{i}\right\}h_{\zeta}$ $(i=1,2,\ldots,N_{\mathrm{coset}},~\zeta=1,2,\ldots,N_h)$ on a position vector $\bm{r}_1$, we can obtain  $i$-th virtual cluster in which 
   $\zeta$-th atom is indexed by $i_{\zeta}$ 
    and the internal coordinates of $i_{\zeta}$-th atom is  
    $\tilde{\bm{\eta}}_{{i}_{\zeta}}=  p_{i} h_{\zeta}
   \bm{r}_1+ \bm{\tau}_{i}$, where $\tilde{\bm{\eta}}_{i_{\zeta}}=  \bm{r}_1$ for $i=1$ and $\zeta=1$~\cite{previous_vc}. 
 For clarity of this construction, 
 virtual  
clusters in space group $P4bm$ are shown in Fig.~\ref{fig:cluster} as an example.  There are two sets of virtual clusters in this case, each of which consists of four atoms in square lattice arrangement since space group $P4bm$ is decomposed as $P4bm=\{E|\bm{0}\}\mathcal{H}\mathcal{T} + \{IC_{2y}|\bm{\tau}\}\mathcal{H}\mathcal{T}$ with $\mathcal{H}=(\{E|\bm{0}\},\{C_{2z}|\bm{0}\},\{C_{4z}|\bm{0}\},\{C^{-1}_{4z}|\bm{0}\})$ and $\bm{\tau}= (\frac{1}{2},\frac{1}{2},0)$.

\begin{figure}
    \centering
    \includegraphics[width=1.0 \hsize]{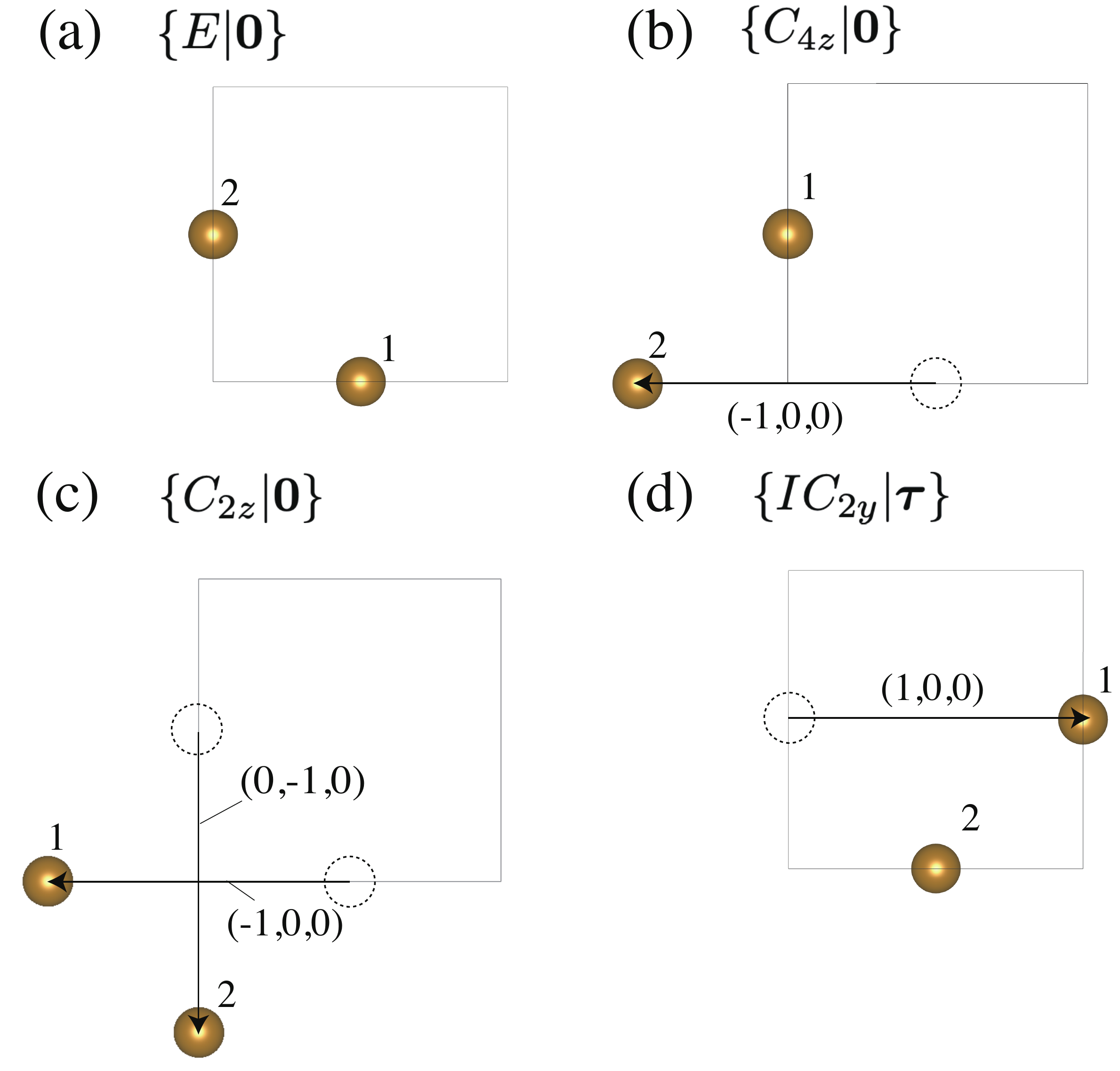}
    \caption{Transformation properties of atoms on $2b$-site in a crystallographic system with space group $P4bm$.  
     The arrow represents primitive translation vector $\bm{R}^{\alpha}_{i_\zeta}$ associated  with each space group operator (see main text).
     }
    \label{fig:trans_crystal}
\end{figure}

To take account of the inter-unit cell spatial modulation characterized by a propagation vector $\bm{k}$, we consider the Fourier transform of magnetic bases as follows: 
\begin{align}
&  \bm{\psi}^{(X)}_{\bm{k} l\Gamma_{\mathrm{s}} \gamma_{\mathrm{s}} i} =  \frac{1}{\sqrt{N}}\sum_{\bm{R}}\bm{\psi}^{(X)}_{\bm{R}l\Gamma_{\mathrm{s}} \gamma_{\mathrm{s}} i}  e^{i\bm{k}\cdot \bm{R}} , \label{eq_magbase_one_vc_k}\\
 & \tilde{\bm{e}}^{\mathrm{axial}}_{\bm{k}i_{\zeta} }  = \frac{1}{\sqrt{N}}\sum_{\bm{R}} \tilde{\bm{e}}^{\mathrm{axial}}_{\bm{R}i_{\zeta} } e^{i\bm{k}\cdot \bm{R}}  , \label{eq_axial_vc_k}
\end{align}
where  
$\tilde{\bm{e}}^{\mathrm{axial}}_{\bm{R}i_{\zeta} }$ 
is the axial unit vector located on the site $i_{\zeta}$ of virtual cluster with position $\tilde{\bm{\eta}}_{i_{\zeta}}+\bm{R}$, and 
$\bm{R}$ is the primitive translation vector assuming that  the virtual clusters are periodically arranged. 
$\bm{\psi}_{\bm{R}l\Gamma_{\mathrm{s}} \gamma_{\mathrm{s}} i}$ are magnetic bases  on the $i$-th virtual cluster  whose centers are located on $\bm{R}+\bm{\tau}_i$ given as follows: 
\begin{align}
 \bm{\psi}^{(X)}_{\bm{R} l \Gamma_{\mathrm{s}} \gamma_{\mathrm{s}} i} 
 &= \sum_{\zeta =1}^{N_h}  \bm{u}^{(X)}_{l \Gamma_{\mathrm{s}} \gamma_{\mathrm{s}} ,  i_{\zeta} } 
 \cdot
 \tilde{\bm{e}}^{\mathrm{axial}}_{ \bm{R} i_{\zeta}},  \label{eq_magbase_one_vc_R}
\end{align}
where $\Gamma_{\mathrm{s}}$ and $\gamma_{\mathrm{s}}$ represent IRREP of point group of a  
virtual cluster and its component, respectively.   
\begin{figure}
    \centering
    \includegraphics[width=1.0 \hsize]{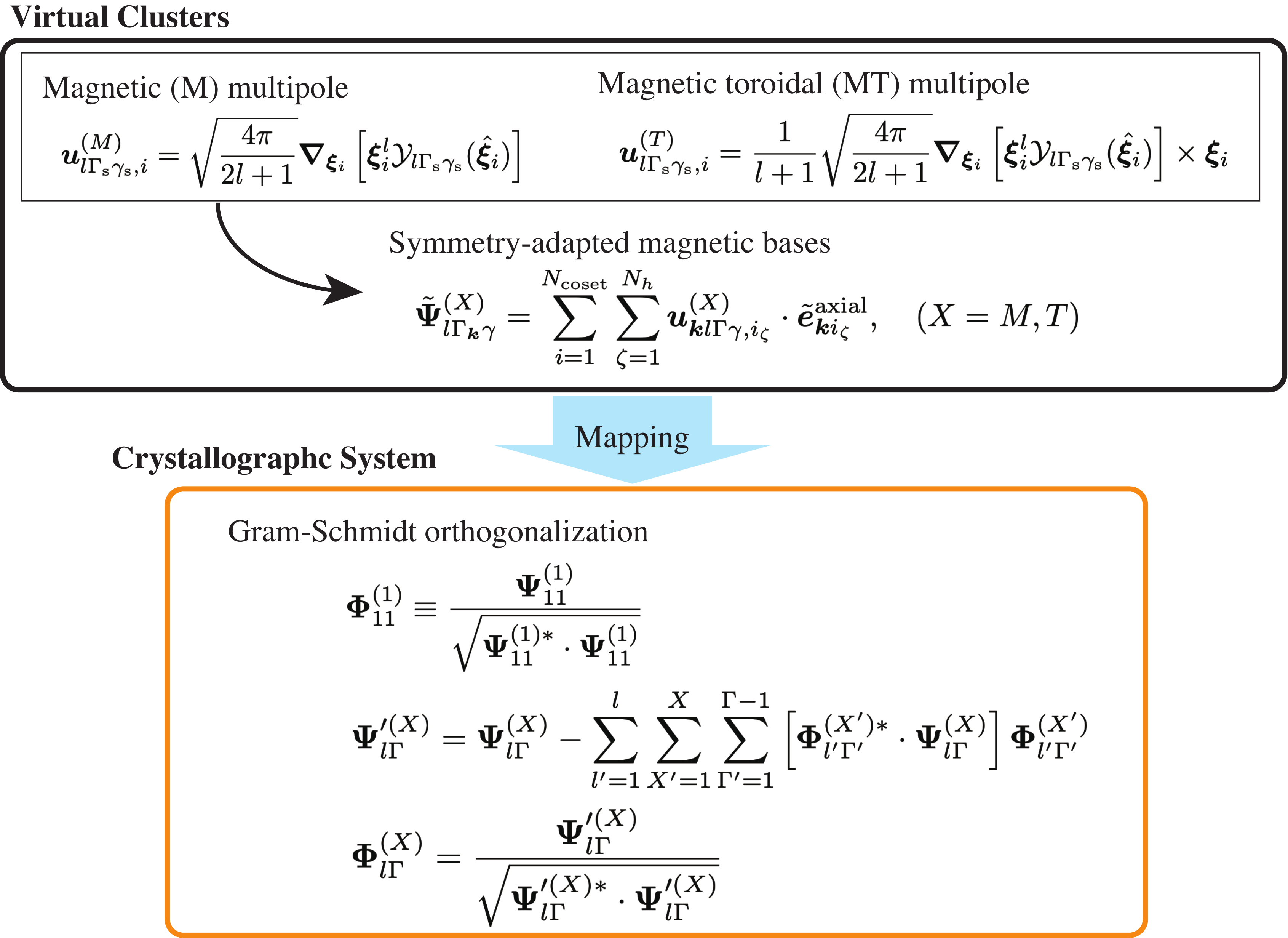}
    \caption{Outline of magnetic bases generation.}
    \label{fig:outline}
\end{figure}
  The coefficient  
  $\bm{u}^{(X)}_{l \Gamma_{\mathrm{s}} \gamma_{\mathrm{s}} ,  i_{\zeta} }$ 
  is obtained by putting 
  $p_ih_{\zeta}\bm{r}_{1}$ 
  into $\bm{\xi}_j$ in  Eqs.~(\ref{eq_M_mp_pg}) and (\ref{eq_T_mp_pg}). 
  We can construct the complex basis vectors having 
   IRREPs 
  under $\bm{k}$-group as linear combinations of the magnetic bases of Eq.~(\ref{eq_magbase_one_vc_k}) as follows:  
\begin{align}
 \tilde{\bm{\Psi}}^{(X)}_{l \Gamma_{\bm{k}}\gamma} 
 &= \sum_{i=1}^{N_{\mathrm{coset}}}  \sum_{\Gamma_{\mathrm{s}}\gamma_{\mathrm{s}}} C^{(X)}_{\bm{k}l \Gamma_{\mathrm{s}} \gamma_{\mathrm{s}} i; \Gamma \gamma}  \bm{\psi}_{\bm{k} l \Gamma_{\mathrm{s}} \gamma_{\mathrm{s}} i}^{(X)}  \nonumber \\
 &= \sum_{i=1}^{N_{\mathrm{coset}}} \sum_{\zeta=1}^{N_h}   
\bm{u}^{(X)}_{l \Gamma_{\bm{k}} \gamma, i_{\zeta}}  
\cdot
 \tilde{\bm{e}}^{\mathrm{axial}}_{\bm{k} i_{\zeta}},  \label{eq_magbase_vc_k}
 \end{align}
where $\Gamma_{\bm{k}}$ denotes the IRREPs of  $\bm{k}$-group, $\gamma$  represents its component, 
 $u^{(X)}_{l \Gamma_{\bm{k}} \gamma,  \alpha \mu} $ 
is given as follows: 
\begin{align}
 & \bm{u}^{(X)}_{l \Gamma_{\bm{k}} \gamma, i_{\zeta} } 
 = \sum_{\Gamma_{\mathrm{s}}\gamma_{\mathrm{s}}} 
 C^{(X)}_{l \Gamma_{\mathrm{s}} \gamma_{\mathrm{s}} i; \Gamma_{\bm{k}} \gamma}  
 \bm{u}^{(X)}_{l \Gamma_{\mathrm{s}} \gamma_{\mathrm{s}}, i_{\zeta} }. \label{eq_uk}
\end{align}
 Note that  coefficients 
 $C_{l \Gamma_{\mathrm{s}}\gamma_{\mathrm{s}} i; \Gamma_{\bm{k}} \gamma}$ 
 can be obtained  
 using ordinary projection operator method to obtain the linear combination of atomic orbitals (LCAO) for the tight-binding model with wave vector $\bm{k}$ and IRREP $\Gamma_{\mathrm{s}}$  at  site $i$~\cite{Kovalev_irrep}. 
 Complex magnetic basis  $\bm{\Psi}^{(X)}_{l \Gamma_{\bm{k}}\gamma} $ in crystallographic systems can be obtained by mapping $\tilde{\bm{\Psi}}^{(X)}_{ l \Gamma_{\bm{k}} \gamma}$ onto crystallographic lattices as follows:   
 \begin{align}
 \bm{\Psi}^{(X)}_{l \Gamma_{\bm{k}}\gamma} 
 &= \sum_{i=1}^{N_{\mathrm{coset}}}  \sum_{\zeta  =1}^{N_h} 
 e^{-i\varphi^{\bm{k},1}_{i_{\zeta}}}
 \bm{u}^{(X)}_{l \Gamma_{\bm{k}} \gamma, i_{\zeta}}
 \cdot
 \bm{e}^{\mathrm{axial}}_{\bm{k} s_{i_{\zeta}}[1]},  \label{eq_magbase_cryst_k}
 \end{align}
 where $\bm{e}^{\mathrm{axial}}_{\bm{k} \alpha}$  represents the Fourier component of axial unit vector on site $\alpha$ in the crystallographic system and  $s_{i_{\zeta}}[\alpha]$ is the permutation operator for indices of atomic sites $\alpha$ associated with the symmetry operation $\{p_ih_{\zeta}|\bm{\tau}_i\}$, which transforms an atomic site of crystallographic system $\bm{\eta}_{\alpha}$  as follows: 
 $\{p_ih_{\zeta}|\bm{\tau}_i\} \bm{\eta}_{\alpha} = \bm{\eta}_{s_{i_{\zeta}}[\alpha]} +  \bm{R}^{\alpha}_{i_{\zeta}} $  with $ \bm{R}^{\alpha}_{i_{\zeta}} $ being a primitive translation vector. $\alpha=1$ represents the atom  mapped from an atom in virtual cluster with $i=1$ and $\zeta=1$. 
Transformation properties of atomic sites on $2b$-site in space group $P4bm$ are shown in Table~\ref{tab:cluster} and Fig.~\ref{fig:trans_crystal} as an example. 
 The phase factor $\varphi^{\bm{k},\alpha}_{i_{\zeta}}$ is given as $ \varphi^{\bm{k},\alpha}_{i_{\zeta}}= \bm{k}\cdot  \bm{R}^{\alpha}_{i_{\zeta}}$. 
  We note that letting $N_{\mathrm{atom}}$ be the number of magnetic atoms in the crystallographic unit cell, each magnetic basis $\bm{\Psi}^{(X)}_{l \Gamma_{\bm{k}}\gamma}$ in Eq.~(\ref{eq_magbase_cryst_k}) is $3N_{\mathrm{atom}}$-dimensional vector while $\tilde{\bm{\Psi}}^{(X)}_{l \Gamma_{\bm{k}}\gamma}$ in Eq.~(\ref{eq_magbase_vc_k}) is $3N_0$-dimensional one with $N_0\ge N_{\mathrm{atom}}$. 
 The magnetic bases in Eq.~(\ref{eq_magbase_cryst_k}) have transformation properties of IRREP $\Gamma_{\bm{k}}$ in $\bm{k}$-group $\mathcal{G}_{\bm{k}}$ as shown in Appendix~A. 
 We can obtain complete orthogonal symmetry-adapted magnetic structure basis set  iteratively using Eqs.~(\ref{eq_M_mp_pg}), (\ref{eq_T_mp_pg}), and (\ref{eq_magbase_one_vc_R})-(\ref{eq_magbase_cryst_k}) combined with the following Gram-Schmidt orthogonalization increasing rank $l$ from $l=1$~\cite{Suzuki_PhysRevB.99.174407},
\begin{align}
&\bm{\Phi}^{(1)}_{11}\equiv \frac{\bm{\Psi}^{(1)}_{11}}{\sqrt{\bm{\Psi}^{({1})*}_{11}\cdot\bm{\Psi}^{(1)}_{11}}},  \\
&\bm{\Psi}'^{(X)}_{l\Gamma}
=
\bm{\Psi}^{(X)}_{l\Gamma}
-
\sum^{l}_{l'=1}\sum^{X}_{X'=1}\sum^{\Gamma-1}_{\Gamma'=1}
\left[\bm{\Phi}^{({X'})*}_{l'\Gamma'}\cdot\bm{\Psi}^{(X)}_{l\Gamma}\right]
\bm{\Phi}^{(X')}_{l'\Gamma'}, \\
&\bm{\Phi}^{(X)}_{l\Gamma}=\frac{\bm{\Psi}'^{(X)}_{l\Gamma}}{\sqrt{\bm{\Psi}'^{({X})*}_{l\Gamma}\cdot\bm{\Psi}'^{(X)}_{l\Gamma}}},
\end{align} 
where $\bm{\Phi}^{(X)}_{l\Gamma}$ is the orthonormal magnetic basis and we reexpress types of multipoles with $X=M$ and $T$ by the numbers $X=1$ and $2$, respectively, and abbreviately denote the IRREP of $\bm{k}$-group $\Gamma_{\bm{k}}$ and its component $\gamma$ by $\Gamma$. The magnetic bases generation procedure is summarized in Fig.~\ref{fig:outline}.
  Although resultant magnetic bases are generally complex, we can always construct real (physical) magnetic bases by taking linear combinations of bases with $\bm{k}$ and $-\bm{k}$. 
 

\subsection{Multiple-$\bm{k}$ states}

\begin{figure*}
    \centering
    \includegraphics[width=1.0 \hsize]{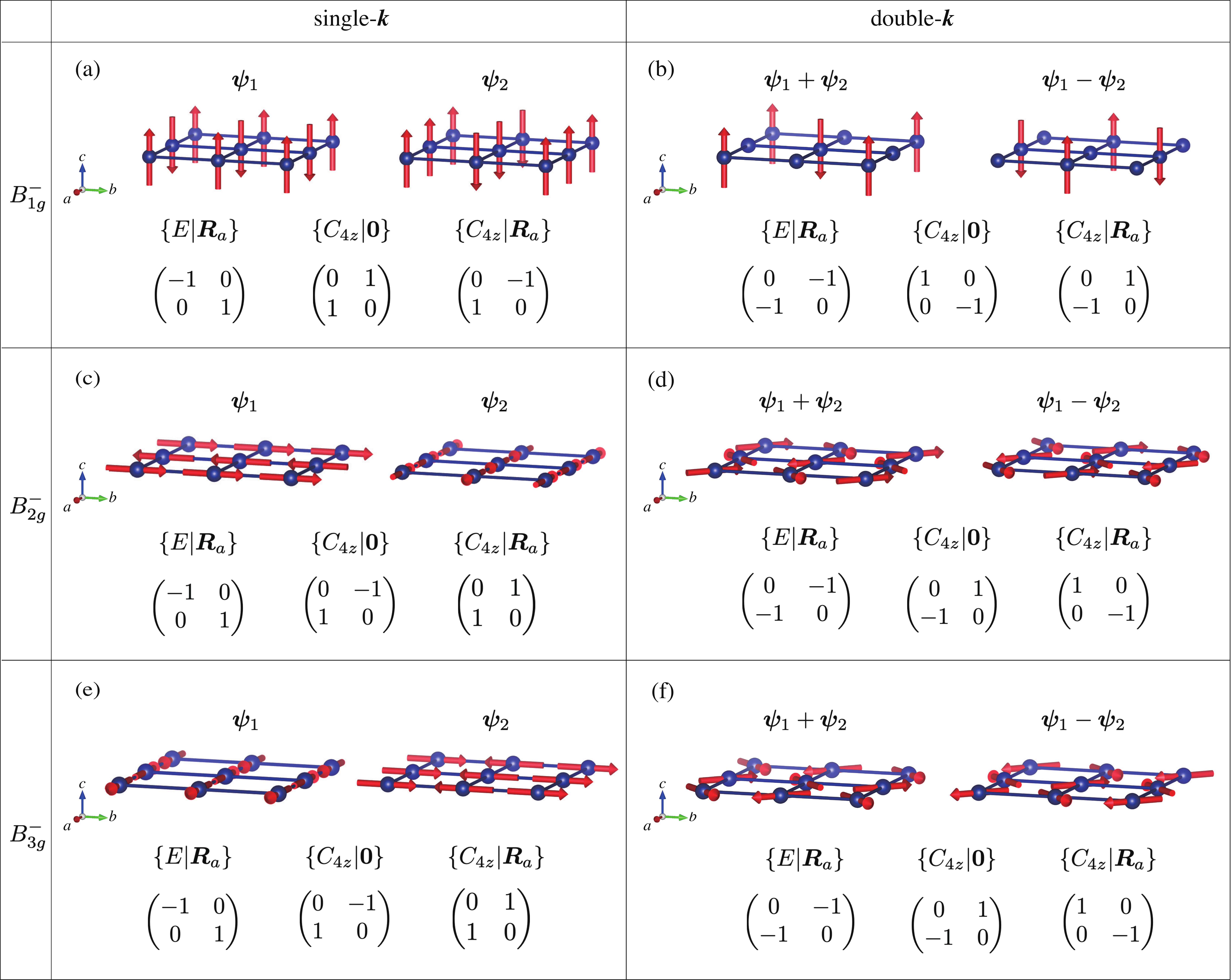}
    \caption{(a), (c), and (e)  Single-$\bm{k}$ and (b), (d), and (f) double-$\bm{k}$ magnetic structures with $\bm{k}=(\frac{1}{2},0,0)$ and $(0,\frac{1}{2},0)$ on the square lattice. 
      The representation matrices for symmetry operations $\{E|\bm{0}\}$, $\{C_{4z}|\bm{0}\}$, and $\{C_{4z}|\bm{R}_a\}$ with $\bm{R}_a=(1,0,0)$ are also shown. 
      } 
    \label{fig:double_q_example}
\end{figure*}
\begin{figure}
    \centering
    \includegraphics[width=1.0 \hsize]{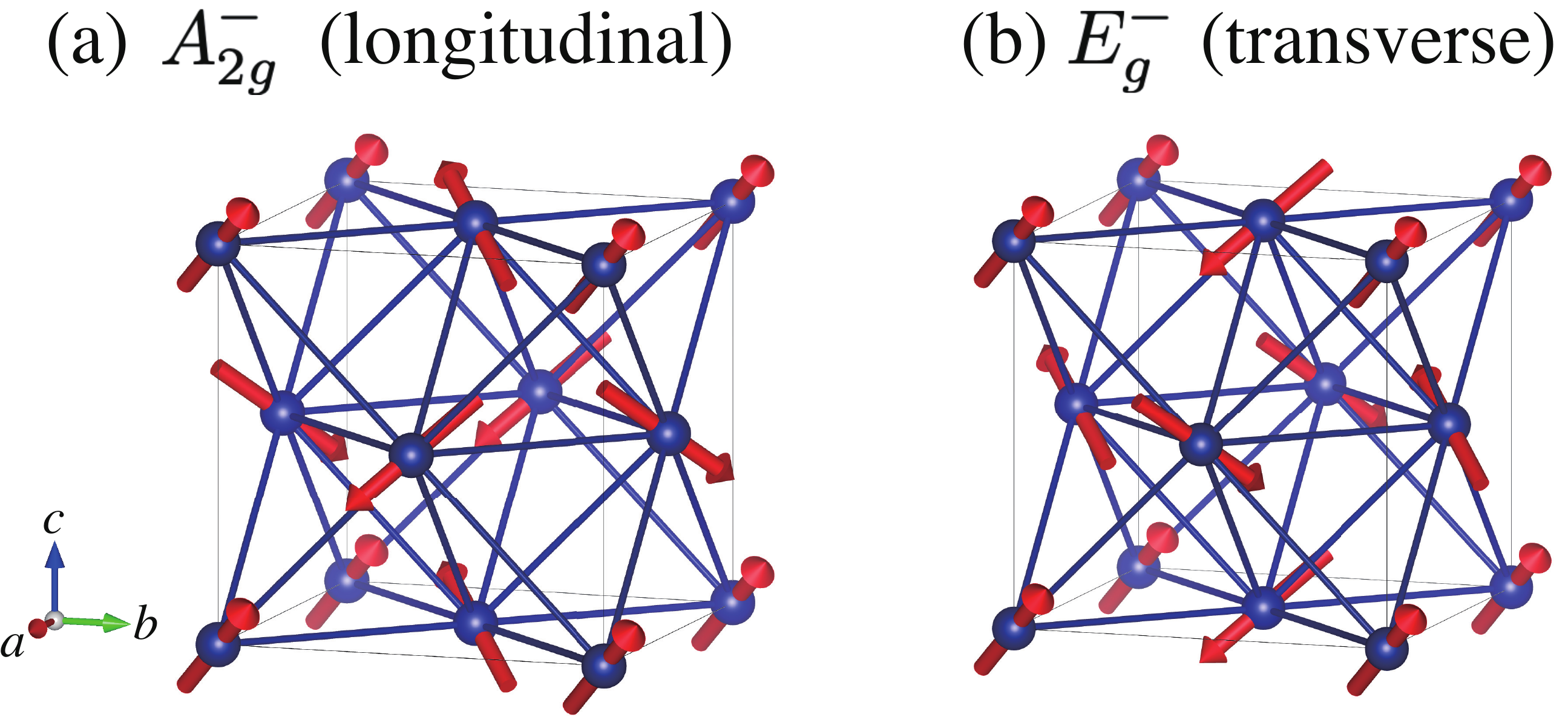}
    \caption{Triple-$\bm{k}$ magnetic structures  on the FCC lattice induced from magnetic structures with $\bm{k}=(1,0,0)$ having (a) $A^-_{2g}$- and (b) $E^-_{g}$-representations.}
    \label{fig:triple_q_example}
\end{figure}
 A single-$\bm{k}$ magnetic structure with propagation vector $\bm{k}$ in an IRREP $\Gamma_{\bm{k}}$ of $\bm{k}$-group $\mathcal{G}_{\bm{k}}$ is a basis vector in  an IRREP of $\mathcal{G}$, that is $(\Gamma_{\bm{k}}\!\uparrow\! \mathcal{G}) \!\downarrow\!\mathcal{G}_{\bm{k}}= \Gamma_{\bm{k}}$~\cite{Bradley_Cracknell_1972,Inui_Group.Theory}.  
  Multiple-$\bm{k}$ magnetic structures are induced as bases of IRREPs in space group $\mathcal{G}$ from single-$\bm{k}$ magnetic structure bases with IRREPs in  $\mathcal{G}_{\bm{k}}$ by taking account of the following relation between $\mathcal{G}_{\bm{k}}$ and  $\mathcal{G}$~\cite{Bradley_Cracknell_1972,Inui_Group.Theory}: 
\begin{align}
\mathcal{G}=\sum_{i=1}^{S_{\bm{k}}}  \left\{p_{i}|\bm{\tau}_{i}\right\}  \mathcal{G}_{\bm{k}},  \label{eq_coset_G_by_Gk}
\end{align}
where  $\left\{p_{1}|\bm{\tau}_{1}\right\}=\left\{E|\bm{0}\right\}$ and $\left\{p_{i}|\bm{\tau}_{i}\right\}$ for $i\ge 2$ is the symmetry operation of space group $\mathcal{G}$ not included in $\bm{k}$-group $\mathcal{G}_{\bm{k}}$. 
The set of wave vectors $\bm{k}_{i}\equiv p_{i}\bm{k}$ 
$(i=1,2,\ldots,S_{\bm{k}})$ 
which are inequivalent 
 with each other are called the star of $\bm{k}$.  
  The bases in IRREP of $\mathcal{G}$ 
 having propagation vector $\bm{k}_{i}$  
 can be 
 generated by acting $\{p_{i}|\bm{\tau}_i\}$  on a single-$\bm{k}$ magnetic structure  with IRREP $\Gamma_{\bm{k}}$ of $\mathcal{G}_{\bm{k}}$.
  Dimension of IRREPs of $\mathcal{G}$ is therefore equal to that of $\mathcal{G}_{\bm{k}}$ multiplied by $S_{\bm{k}}$. 
 The multiple-$\bm{k}$ magnetic structures formed as linear combination of the  symmetry-adapted single-$\bm{k}$ structures with propagation vectors associated with each other by symmetry operations $\left\{p_{i}|\bm{\tau}_{i}\right\}$ in Eq.~(\ref{eq_coset_G_by_Gk}) 
  can  
  preserve the higher rotational symmetry than that of the single-$\bm{k}$ structures, while it makes the translational symmetry lower than that under the single-$\bm{k}$ structures.

To gain a clear insight into multiple-$\bm{k}$ magnetic states, 
 we here discuss two examples of multiple-$\bm{k}$ states: double-$\bm{k}$ states on the square lattice and triple-$\bm{k}$ states on the face centered cubic (FCC) structure. 
 The double-$\bm{k}$ states having high magnetic point group symmetry with propagation vectors $\bm{k}=(\frac{1}{2},0,0)$ and $(0,\frac{1}{2},0)$ on the square lattice as well as the single-$\bm{k}$ states are shown in   
 Fig.~\ref{fig:double_q_example}. 
 The double $\bm{k}$-states are obtained as linear combinations of single-$\bm{k}$ states $\bm{\psi}_1$ and $\bm{\psi}_2$ as shown in Fig.~\ref{fig:double_q_example}. 
 Note that the space group of the square lattice is $P4/mmm$ and 
 point group preserving  $\bm{k}$, referred as $\bm{k}$-point group, 
 is $D_{2h}$.  
 In this case, all IRREPs of $\mathcal{G}_{\bm{k}}$ are one-dimensional and $S_{\bm{k}}=2$, leading to two-dimensional IRREPs under $\mathcal{G}$. 
 The representation matrices for several space group elements are also explicitly shown in Fig.~\ref{fig:double_q_example}. 
 We can see that in the  single-$\bm{k}$ (double-$\bm{k}$) states, the representation matrix  for translational symmetry $\{E|\bm{R}_a\}$ with $\bm{R}_a=(1,0,0)$ has diagonal (non-diagonal) form, while  that for rotational operation $\{C_{4z}|\bm{0}\}$ or  $\{C_{4z}|\bm{R}_a\}$ has non-diagonal (diagonal) form. 
  This indicates that the rotational symmetry in 
 double-$\bm{k}$ states is higher than that in 
 the single-$\bm{k}$ states but    
 translational symmetry in the double-$\bm{k}$ states is lower than that in the single-$\bm{k}$ states, as mentioned before.  
  It should be noted that time-reversal symmetry is retained macroscopically both in the single- and double-$\bm{k}$ states, that is,  time-reversal operation combined with translational operation is preserved in the magnetic states. 
 As a result,  the magnetic point group in the double-$\bm{k}$ states shown in Fig.~\ref{fig:double_q_example} is same as in the paramagnetic state, that is $4/mmm1'$, whereas that in the single-$\bm{k}$ states is 
 orthorhombic one $mmm1'$, which is lower than  $4/mmm1'$.  
 The breaking of four fold rotational symmetry along $z$-axis in the single-$\bm{k}$ states is 
  understood from the uniaxial  folding of the Brillouin zone  
 along $x$- or $y$-direction from the paramagnetic state to magnetic state. 
  This type of magnetic ordering causing breaking of four fold rotational symmetry has been discussed in the context of iron-pnictides~\cite{Kubo_JPSJ.78.083704,Lee_PhysRevLett.103.267001}. 

 
The examples of triple-$\bm{k}$ states with propagation vectors at the X-points, $\bm{k}_1=(1,0,0)$, $\bm{k}_2=(0,1,0)$, and  $\bm{k}_3=(0,0,1)$,  on FCC lattice are also shown in   Figs.~\ref{fig:triple_q_example}(a) and (b). 
The space group of FCC lattice is  $Fm\bar{3}m$ and the $\bm{k}$-point group is $D_{4h}$. 
Note that the magnetic structure in Fig.~\ref{fig:triple_q_example}(a) has been observed in Fe-Mn   alloys~\cite{KOUVEL1963529,Umebayashi_JPSJ.21.1281,Endoh_JPSJ.30.1614} and several $f$-electron compounds such as Np$X$ ($X$=As, Sb, Bi)~\cite{BURLET1987151,SANCHEZ1988999,BURLET1992131} and  USb~\cite{ROSSATMIGNOD1980237,LANDER19957}.  
 In contrast to 
  the square lattice discussed above,  
 there are two types of two-dimensional IRREPs, $E^-_{g}$- and $E^-_{u}$-IRREPs, in addition to the eight types of one-dimensional IRREPs of $D_{4h}$ $\bm{k}$-point group at the X-points.  
Here and hereafter, we label the odd (even)-parity for time-reversal symmetry as superscript $-$ ($+$) in IRREPs such as $E^{-}_{g}$ ($E^{+}_{g}$).   
 The two- and one-dimensional IRREPs under $\mathcal{G}_{\bm{k}}$ with $S_{\bm{k}}=3$ form six- and three-dimensional IRREPs under $\mathcal{G}$ since the X-points $(1,0,0)$, $(0,1,0)$, and $(0,0,1)$ are transformed into each other by symmetry operations $\{E|\bm{0}\}$, $\{C_{3[111]}|\bm{0}\}$, and  $\{C^2_{3[111]}|\bm{0}\}$, where $C_{3[111]}$ is three fold rotation along $[111]$ direction. 
 The single-$\bm{k}$ magnetic state with propagation vector $\bm{k}_i$ ($i=1,2,3$) preserves 
 macroscopic time-reversal symmetry 
  since $\bm{k}_i$ is a time-reversal invariant momentum.
   On the contrary, 
   any linear combination of single-$\bm{k}$ magnetic bases with $\bm{k}_i$ to form  triple-$\bm{k}$ states necessarily breaks  macroscopic time-reversal symmetry. 
   For instance, 
     when a translational operation $\bm{t}$ carries the factor $-1$ for the magnetic bases with $\bm{k}_1$ and $\bm{k}_2$, and then same translation carries factor $+1$ for that with $\bm{k}_3=-\bm{k}_1-\bm{k}_2$. 
     Since all the magnetic bases with $\bm{k}_i$ ($i=1,2,3$) cannot be simultaneously transformed into themselves with factor $-1$ under a primitive translation $\bm{t}$, the triple-$\bm{k}$ magnetic ordering  breaks the macroscopic time-reversal symmetry. 
  As a result, the triple-$\bm{k}$ magnetic order is accompanied with the secondary uniform magnetic order parameter, that is, magnetic multipoles with $\bm{k}=\bm{0}$.  
  In the present case, triple-$\bm{k}$ magnetic ordering induces the secondary uniform magnetic order parameter  having $A^-_{2g}$-IRREP of $O_h$ point group. 
   The  experimentally  observed  all-in-all-out type magnetic order in a pyrochlore oxide Cd$_2$Os$_2$O$_7$ with space group $Fd\bar{3}m$~\cite{Yamaura_PhysRevLett.108.247205}  has the same symmetry of the secondary magnetic order parameter, $A^-_{2g}$, of the triple-$\bm{k}$ state in the FCC lattice, as shown in Fig.~\ref{fig:triple_q_example}(a).  
 Arima has pointed out that the all-in-all-out magnetic order can be regarded as a ferroic order of magnetic octupole $M_{xyz}$ and  leads to several intriguing phenomena such as asymmetric magnetization and   linear magnetostriction~\cite{Arima_JPSJ.82.013705}.
  It is, therefore, expected that similar phenomena emerge also in Fe-Mn alloys, Np$X$, and USb. On the other hand, the triple-$\bm{k}$ state shown in  Fig.~\ref{fig:triple_q_example}(b)  is induced from the magnetic structure with $E^-_g$-symmetry of $\bm{k}$-group at X-point. 
 This type of magnetic ordering  has been observed in  UO$_2$~\cite{BURLET1986121,Lander_2020}. 
 This magnetic ordering with magnetic point group $m\bar{3}$ is absence of anti-unitary symmetry operations, which  belongs to type-I Shubnikov group~\cite{Bradley_Cracknell_1972}.
   The absence of anti-unitary symmetry operations indicates that 
   the axial (polar) tensors having opposite parities for time-reversal operation, that is, M- (MT-) and ET- (E-) multipoles cannot be distinguished from the symmetry viewpoint.    These multipoles thus are simultaneously  activated in this situation~\cite{Yatsurhiro_PhysRevB.104.054412}, resulting in intriguing physical phenomena. 
  For instance, electric toroidal dipolar moment as well as the magnetic moment are linearly coupled with magnetic field in the magnetic state shown in Fig.~\ref{fig:triple_q_example}(b). 
  The crystal symmetry is thus lowered to $C_{2h}$ by magnetic field along $[001]$-axis in contrast to the case of similar paramagnetic point group $m\bar{3}1'$, in which the crystal symmetry is lowered to $D_{2h}$ by magnetic field along $[001]$-axis.  
  Such characteristic magnetostriction is useful for identifying the order parameter. 


\section{Application to representative AHE antiferromagnets}
\subsection{$\alpha$-Mn}

\begin{figure}
    \centering
    \includegraphics[width=1.0 \hsize]{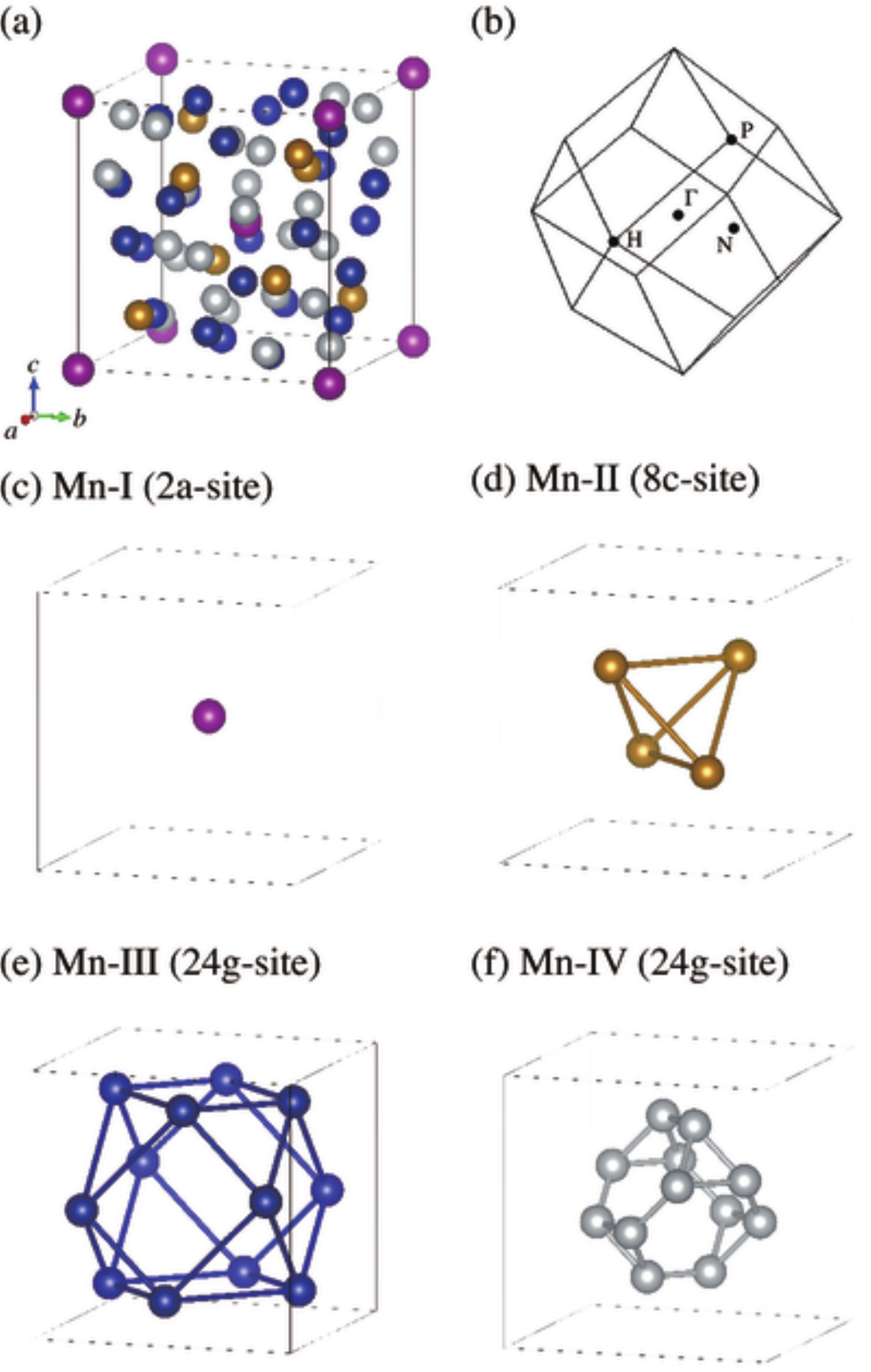}
    \caption{(a) Crystal structure and (b) first Brillouin zone of $\alpha$-Mn. Mn-I, -II, -III, and -IV sites are separately shown in panels (c), (d), (e), and (f), respectively, where body-centered sites are not shown.}
    \label{fig:cryst_aMn}
\end{figure}

Unary manganese forms various crystal structures depending on temperature, which are called $\alpha$-, $\beta$-, $\gamma$-, and $\delta$-Mn~\cite{Bigdeli_pssb.201552203}. 
   $\alpha$-Mn with body centered cubic (BCC) structure, 
  which is realized 
  at $T\lesssim \SI{980}{\kelvin}$, has attracted interest due to its intriguing and complex magnetic properties. 
 A noncollinear antiferromagnetism with ordering vector $\bm{k}= \bm{k}_{\mathrm{H}}\equiv (1,0,0)$ is realized at   $T\lesssim \SI{95}{\kelvin}$~\cite{Shull_RevModPhys.25.100,Yamada_JPSJ.28.596,Yamada_JPSJ.28.615,Yamagata_JPSJ.33.400,Lawson_JApplPhys.76.7049} under an ambient pressure, while  a 
  first order transition to the another magnetic phase with tiny net magnetization  $\sim 0.02~\mu_{\mathrm{B}}/\mathrm{Mn}$ 
   takes place under an applied pressure $P\sim \SI{1.4}{\giga\pascal}$~\cite{Takeda_JPSJ.77.025001,Akiba_PhysRevResearch.2.043090,Ito_JPSJ.90.085001}.  
 The recent experimental study has revealed the emergence of large anomalous Hall response reaching $\sigma_{xy}\sim 400$-$\SI{600}{\siemens\per\centi\meter}$ in the high pressure phase despite such small net magnetization~\cite{Akiba_PhysRevResearch.2.043090}.
    We here apply our theory to $\alpha$-Mn to generate magnetic structures at high symmetry $\bm{k}$-points and provide a possible scenario for large anomalous Hall response   
   in the high pressure magnetic phase, in which 
 magnetic structure has not been clarified experimentally.
\tabcolsep = 7pt
\begin{table}[b]
    \centering
    \caption{Local magnetic moments in units of $\mu_{\mathrm{B}}$ on each Mn site in AFM phase in $\alpha$-Mn at ambient pressure 
    from Ref.~\cite{Lawson_JApplPhys.76.7049}. 
     Note that Mn-III and Mn-IV sites are split into two different types of Mn sites due to lacking of $C_3$ symmetry in AFM phase. }
    \label{tab:mom_aMn}
    \begin{tabular}{ccccccccc}
    \hline
    \hline
     & I & II & III-1 & III-2 & IV-1 & IV-2 \\   \hline
    $m_x$   & $0.0$ & $0.14$ & $0.43$ & $-0.25$ & $0.27$ & $-0.08$ \\
    $m_y$   & $0.0$ & $0.14$ & $0.43$ & $-0.25$ & $0.27$ & $-0.45$ \\
    $m_z$ & $2.83$ & $1.82$ & $0.43$ & $-0.32$ & $-0.45$& $-0.48$ \\
    $|\bm{m}|$ & $2.83$ & $1.83$ & $0.74$ & $0.48$ & $0.59$& $0.66$ \\
    \hline
    \hline
    \end{tabular}
\end{table}

    \begin{figure*}
    \centering
    \includegraphics[width=1.0 \hsize]{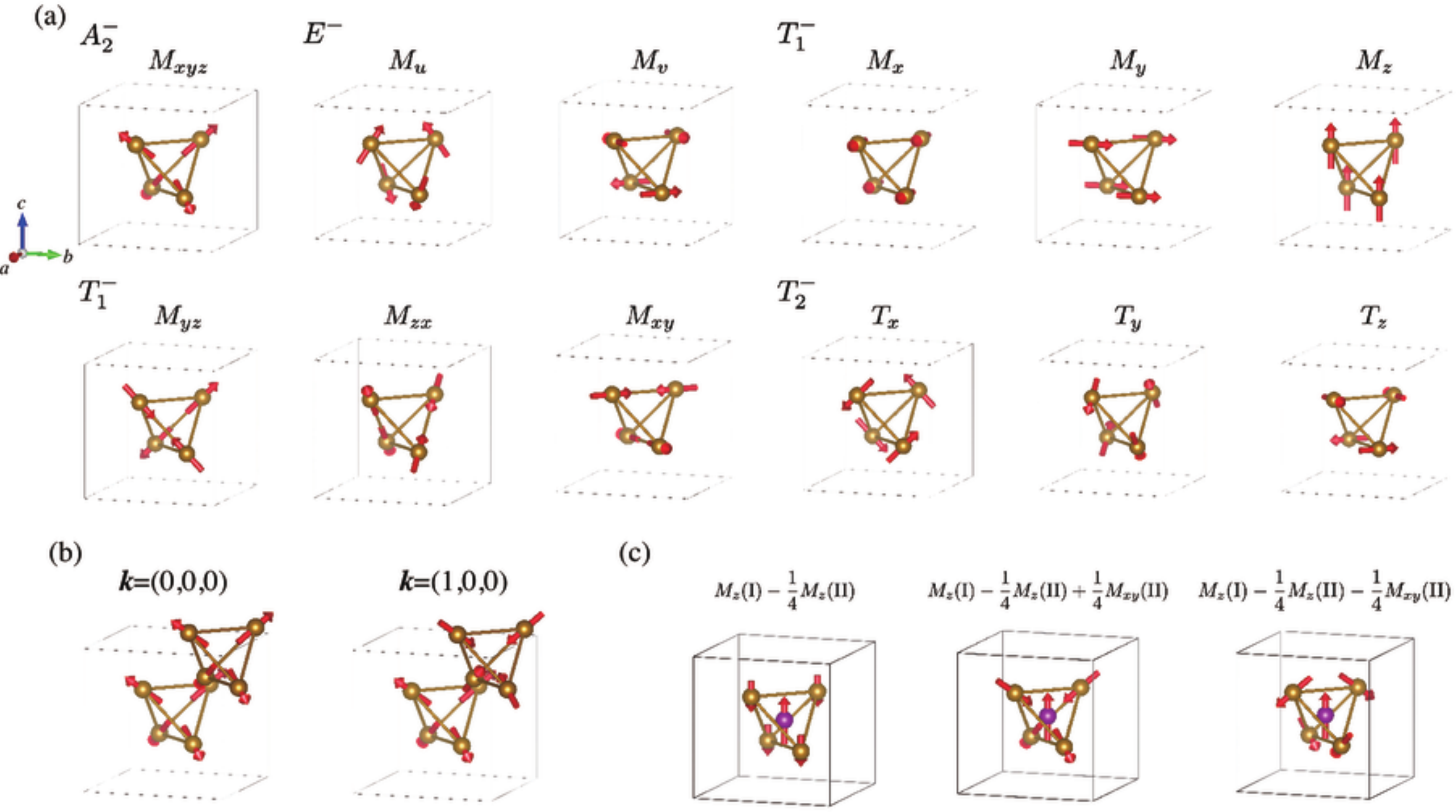}
    \caption{(a) Magnetic structure bases with $\bm{k}=\bm{0}$ on Mn-II sites in $\alpha$-Mn.  (b) Comparison between magnetic basis having $A^{-}_2$-IRREP with $\bm{k}=\bm{0}$ and that with $\bm{k}=(1,0,0)$.
     (c) Examples of magnetic structures  without net magnetization compatible with finite AHE $\sigma_{xy}\neq 0$. 
    }
    \label{fig:mag_q0_aMn}
\end{figure*}

 Figure~\ref{fig:cryst_aMn} shows crystal structure and the corresponding first Brillouin zone for $\alpha$-Mn. The  primitive (conventional) unit cell contains 29 (58) Mn-atoms and resulting crystal structure is relatively complicated.  The space group symmetry of $\alpha$-Mn is $I\bar{4}3m$ (No. 217, $T^{3}_d$), and there are four distinct types of Mn-sites,  Mn-I, Mn-II, Mn-III, and Mn-IV sites which are located on $2a$, $8c$, $24g$, and $24g$ Wyckoff positions, respectively.   The sizes of local magnetic moments largely depend on the types of Mn-sites according to experiments in the AFM phase at ambient pressure. 
  The magnetic moments on Mn-I and -II sites are quite larger than those on  Mn-III and -IV sites, as shown in Table~\ref{tab:mom_aMn}~\cite{Yamada_JPSJ.28.615,Yamagata_JPSJ.33.400,Lawson_JApplPhys.76.7049} 
    and a similar tendency for local magnetic moments is obtained by the first-principles calculations~\cite{Hobbs_jphys_2001,Hobbs_PhysRevB.68.014407,Ehteshami_PhysRevMaterials.1.073803,Pulkkinen_PhysRevB.101.075115}. 
  In addition, the internal magnetic fields on Mn-III and -IV sites are rapidly suppressed by an applied pressure according to the recent zero-field NMR measurements~\cite{Fukazwa_nmr_pc}. 
    We thus  focus on magnetic structures on Mn-I and -II sites in this paper.

 Nonzero anomalous Hall response requires symmetry breaking same as ferromagnetic ordering, which belongs to  $T^{-}_1$-IRREP with $\bm{k}=\bm{0}$ under $T_d$-point group in the present system~\cite{Suzuki_PhysRevB.95.094406,Grimmer:du0368,Martinez_PhysRevB.92.085138,Watanabe_PhysRevB.98.245129,Hayami_PhysRevB.98.165110,Yatsurhiro_PhysRevB.104.054412}.  
  According to the recent neutron diffraction measurements, the magnetic reflection observed in the ambient pressure phase is absent in the high pressure phase~\cite{Kobayashi_neutron_pc}. Therefore, we focus on the cases of magnetic orderings having $T^{-}_1$ symmetry with ordering vector $\bm{k}=\bm{0}$ and $\bm{k}=\bm{k}_{\mathrm{H}}=(1,0,0)$ in the present study.

  The transformation properties for arbitrary magnetic structures on each Wyckoff position  are encoded in representation matrices, which are called magnetic representations. 
  The magnetic representations of $2a$ and $8c$-sites, $D^{(\mathrm{mag})}_{2a}$ and $D^{(\mathrm{mag})}_{8c}$,  are decomposed into IRREPs of $\mathcal{G}_{\bm{k}}$ with $\bm{k}=\bm{0}$ as follows:
  \begin{align}
& D^{(\mathrm{mag})}_{2a} = T^{-}_{1}, \label{eq_mag_aMn_2a}\\
& D^{(\mathrm{mag})}_{8c} = A^{-}_{2}\oplus E^{-} \oplus 2T^{-}_{1} \oplus T^{-}_{2},\label{eq_mag_aMn_8c}
\end{align}
 where we use Mulliken notation for IRREPs under $T_d$ point group since IRREPs of $\mathcal{G}_{\bm{k}}$ with $\bm{k}=\bm{0}$ are equivalent to those of crystallographic point group of $\mathcal{G}$. 
 We show the symmetry-adapted magnetic structures at $8c$-site with $\bm{k}=\bm{0}$ in Fig.~\ref{fig:mag_q0_aMn}(a). 
 Note that the magnetic structures having $\bm{k}=\bm{k}_{\mathrm{H}}$ can be obtained by reversing magnetic moments on body center tetrahedra for those with $\bm{k}=\bm{0}$ as shown in Fig.~\ref{fig:mag_q0_aMn}(b).   The obtained magnetic structure bases with $\bm{k}=\bm{k}_{\mathrm{H}}$ are consistent with the preceding group theoretical analysis of neutron diffraction experiment~\cite{Yamada_JPSJ.28.596}. 
 
 From Eqs.~(\ref{eq_mag_aMn_2a}) and (\ref{eq_mag_aMn_8c}) and Fig.~\ref{fig:mag_q0_aMn}(a),  we can see that there are two sets of magnetic structure bases on $8c$-site with $T^{-}_1$-IRREP  whereas there is one set of magnetic structures on $2a$-site  with $T^{-}_1$-IRREP, that is, ferromagnetic structure. 
  One of $T^{-}_1$-IRREP magnetic structures on $8c$-site corresponds to the ordinary ferromagnetism, and the other  is  noncollinear AFM without net magnetization.  
  The  linear combinations of these magnetic bases can generate nonzero anomalous Hall response $\sigma_{xy}\neq 0$. 
 The magnetic structures without net magnetization are shown in  Fig.~\ref{fig:mag_q0_aMn}(c).

\begin{table*}
    \centering
    \caption{Band deformation and macroscopic responses in the magnetic ordered phases having $T^-_1$-IRREP with the ordering vectors $\bm{k}=\bm{0}$ ($\Gamma$-point) in the high pressure phase and $\bm{k}=\bm{k}_{\mathrm{H}}$ ($\mathrm{H}$-point) in the ambient pressure phase. AHE, ANE, ME, MC, and MOKE represent  anomalous Hall, anomalous Nernst, magneto-electric, magneto-current, and magneto-optical Kerr effects, respectively. 
     Note that magneto-current effects correspond to Edelstein effect and its inverse response.}
    \begin{tabular}{c c c c c c}
    \hline\hline
    ordering vector & \multicolumn{4}{c}{band deformation} & macroscopic response \\\hline
    $\Gamma : (0,0,0)$ & $(3k^2_z-k^2)\sigma_0 $ & $k_x\sigma_x-k_y\sigma_y$  & $\sigma_z$  
                                                                                                                                                           & $ k_z(k_x^2-k_y^2)\sigma_0 $  & AHE, ANE, MOKE, 
                                                                                                                                                           ME, MC \\ 
    $\mathrm{H}: (1,0,0)$ & $(3k^2_z-k^2)\sigma_0$ & $k_x\sigma_x-k_y\sigma_y$ & \ding{55} &\ding{55} & MC \\ \hline\hline
    \end{tabular}
    \label{tab:response_aMn}
\end{table*}

 We here discuss the secondary order parameters based on the Landau theory~\cite{Toledano_Landau.Theory,Lyubarskii_Landau.Theory}, which gives clear insight into physical phenomena in the ordered phases and useful information about primary order parameters. In fact, detailed analysis for couplings between primary and secondary order parameters leads to deep understanding of complex multipolar ordered phases in the $f$-electron systems~\cite{Kusunose_JPSJ.77.064710,Shiina_JPSJ.66.1741,Shiina_JPSJ.67.941,Kuramoto_JPSJ.78.072001,Suzuki_PhysRevB.90.184407}. 
 The possible secondary order parameters are  characterized by $\bm{k}=\bm{0}$ in case the primary order parameter has the ordering vector $\bm{k}=\bm{0}$. 
  We first discuss the electric degrees of freedom induced by the magnetic orderings, which are relevant to elastic response and Edelstein effect, for instance. 
  The electric order parameters, which are time-reversal-even, are induced by the coupling of the electric multipoles to the even order terms of magnetic ones in Landau free energy expression due to the time-reversal symmetry in the paramagnetic phase. 
 The electric order parameters emerging through the lowest third order terms,  can be deduced by irreducible decomposition of symmetric product of primary order parameters as follows:
 \begin{align}
 [T^{-}_1\otimes T^{-}_1] =   A^{+}_{1} \oplus  E^{+} \oplus T^{+}_{2}, 
 \end{align}
 where the secondary order parameters with $A^{+}_{1}$,  $E^{+}$, and  $T^{+}_{2}$-IRREPs correspond to 
 uniform electric monopole $Q_0$, electric quadrupoles $(Q_{u(=3z^2-r^2)},Q_{v(=x^2-y^2)})$, and $(Q_{yz},Q_{zx},Q_{xy})$, respectively from the symmetry viewpoint. 
 Note that in the present case, electric toroidal quadrupole (electric dipole) $(G_{v},G_{u})$ [$(Q_x,Q_y,Q_z)$] also emerges when electric quadrupole $(Q_{u},Q_{v})$ [$(Q_{yz},Q_{zx},Q_{xy}$)] is finite since those have the same IRREP $E^+$ ($T^+_2$) under $T_d$ point group.    The explicit forms of the third order terms in Landau free energy are given as follows: 
 \begin{align}
    F^{(3)} & 
     = c_1\left[M_{y} M_{z}Q_{yz}+(\mathrm{cyclic~perm.})\right] \nonumber \\
    &+c_2 \sqrt{3}\left(M^2_{x}-M^2_{y}\right) Q_{v}\nonumber \\
    &+c_2\left(2M^2_{z}-M^2_{x}-M^2_{y}\right) Q_{u}, \label{eq_3rd_Landau_aMn}
 \end{align}
 where we neglect trivial coupling with the fully symmetric $Q_0$ term 
 and $(\mathrm{cyclic~perm.}) $ represents the terms obtained by cyclic permutations for indices of $M_yM_zQ_{yz}$ as $x,y,z \to y,z,x$ and $z,x,y$, and $c_1$ and $c_2$ are coupling constants determined from microscopic models. 
 Eq.~(\ref{eq_3rd_Landau_aMn}) represents that the electric quadrupole is inevitably induced by primary order parameter with $T^-_1$-IRREP. For example, $Q_{u}$-type electric quadrupole emerges in the case that $M_x=M_y=0$ and $M_z\neq 0$   while  the electric quadrupoles having $T^+_2$ symmetry emerge with the same amplitude of all components, that is,  $Q_{yz}=Q_{zx}=Q_{xy} \neq 0$ in the case that $M_x=M_y=M_z\neq 0$.  Therefore, the symmetry of crystal structure is lowered to 
 tetragonal $D_{2d}$ point group in the former case, which can induce gyrotropic magnetic effect, and polar trigonal $C_{3v}$ point group in the latter case.  
  Note that   
 similar couplings to electric multipoles with $\bm{k}=\bm{0}$ emerges also in the case of the primary order parameter with $\bm{k}=\bm{k}_{\mathrm{H}}$ 
 since  
 the even powers of primary order parameters carry the wave vector $\bm{k}=\bm{0}$ modulo reciprocal lattice vector.

  In the similar manner, secondary magnetic order parameters are discussed. Due to the time reversal symmetry in the nonmagnetic phase, the  secondary magnetic order parameter emerges through the fourth order coupling in the Landau free energy, which can be obtained by the irreducible decomposition of fully symmetric product $[T^{-}_1\otimes T^{-}_1 \otimes T^{-}_1]$ as follows: 
  \begin{align}
 [T^{-}_1\otimes T^{-}_1 \otimes T^{-}_1] =   A^{-}_{2} \oplus  2T^{-}_1 \oplus T^{-}_{2}, 
 \end{align}
where $A^-_{2}$ and $T^{-}_2$ correspond to magnetic octupoles $M_{xyz}$ and $(M^{\beta}_x,M^{\beta}_y,M^{\beta}_z)$ from the symmetry viewpoint. 
 The fourth order term of Landau free energy which represent coupling of order parameter with $T^{-}_1$-IRREP to the secondary order parameters having different symmetry is given as follows:
\begin{align}
    F^{(4)} & 
 = 
    d_1 M_xM_yM_zM_{xyz}  \nonumber \\
    & + d_2 \left[\left(M^2_y-M^2_z\right)M_xM^{\beta}_x+(\mathrm{cyclic~perm.})\right].  
\end{align}
From the above equation, we can see that in the case of $M_x=M_y=0$ and $M_z\neq 0$, the magnetic secondary order parameter does not emerge while in the case of $M_x=M_y=M_z\neq 0$ ($M_z=0$ and $M_x=M_y\neq 0$), the magnetic octupole $M_{xyz}$ ($M^{\beta}_{x}=-M^{\beta}_{y}$) is induced.  On the other hand, when the primary order parameter has a modulation vector $\bm{k}=\bm{k}_{\mathrm{H}}$, secondary magnetic order parameters with $\bm{k}=\bm{0}$ do not emerge since combined symmetry of time-reversal operation and primitive translation retains in the ordered phase.  The secondary induced magnetic multipoles have ordering vector $\bm{k}=\bm{k}_{\mathrm{H}}$ in this case. 

The primary order parameters with inducing the secondary order parameters discussed above give rise to deformation of electronic states and resulting physical phenomena in the ordered phases.  
In particular, the order parameters with the ordering vector $\bm{k}=\bm{0}$ are important since those can be sources of macroscopic responses  according to Neumann's principle~\cite{Birss_symmetry,Grimmer:du0368,Watanabe_PhysRevB.98.245129,Hayami_PhysRevB.98.165110} 
   and experiments using macroscopic probes can detect the order parameters relatively easier.  
  The magnetic order parameter $M_z$ with $\bm{k}=\bm{0}$ is finite in the high pressure phase, while that vanishes in the ambient pressure phase.  
 The former case   
 leads to anomalous Hall, Nernst effects, and magneto-optical  effects.  
On the other hand, electric 
  order parameters 
 $Q_{u}$ and $G_{v}$ with $\bm{k}=\bm{0}$ can be finite in both ambient and high pressure magnetic phases. This gives rise to the antisymmetric spin-splitting of band structures $k_x\sigma_x-k_y\sigma_y$, resulting in  natural optical activity and magneto-current effect such as Edelstein effect~\cite{Levitov_JETP.61.133,EDELSTEIN1990233,Zhong_PhysRevLett.116.077201}, where $\sigma_i$ ($i=x,y,z$) is the Pauli spin matrix. 
We summarize band deformations 
 as the hopping terms in the $2\times 2$ Hamiltonian with spin degree of freedom 
 and macroscopic responses in the ordered phases in Table~\ref{tab:response_aMn}. 

\subsection{Co$M_3$S$_6$}

\begin{figure}[t]
    \centering
    \includegraphics[width=1.0 \hsize]{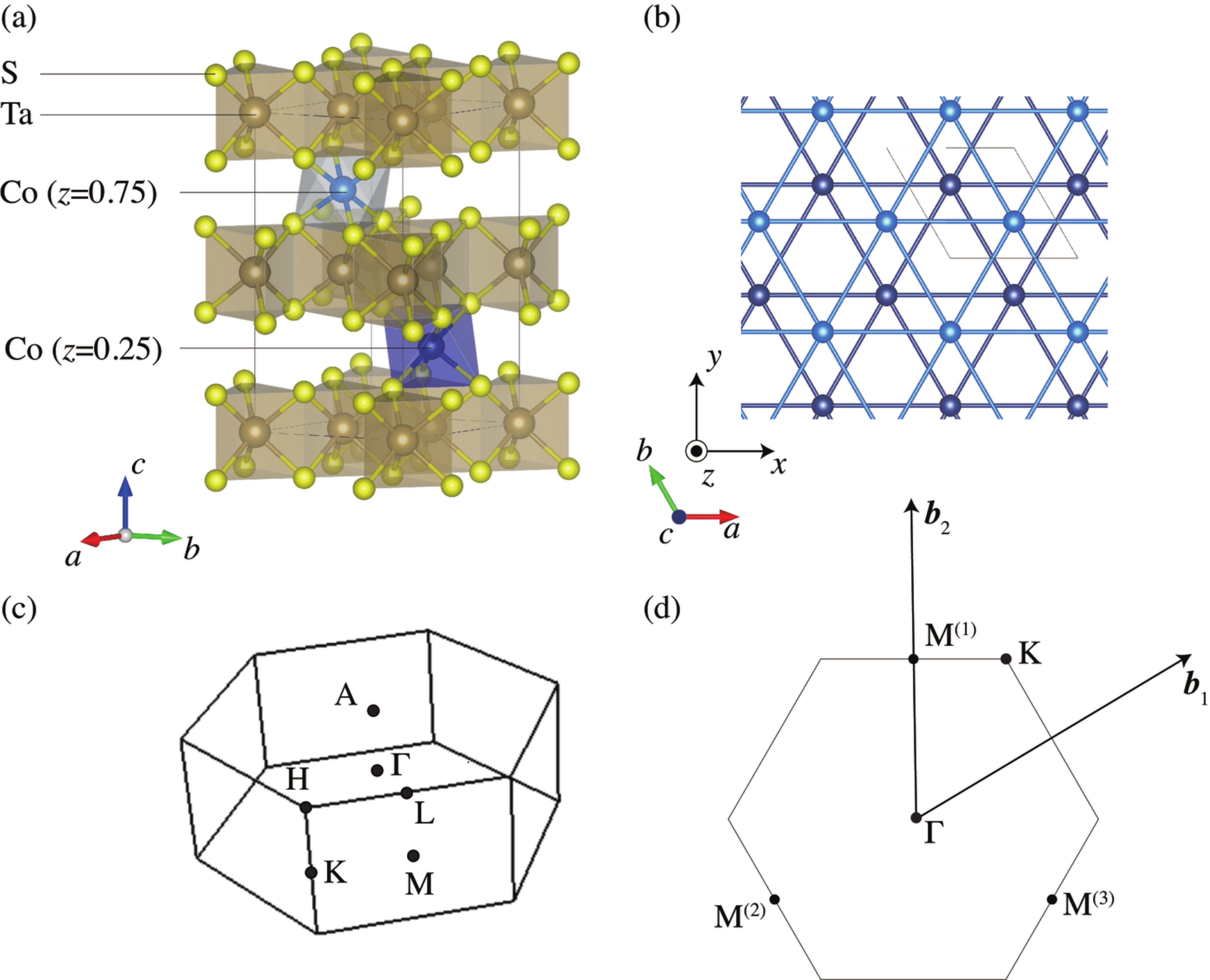}
    \caption{(a) Crystal structure of CoTa$_3$S$_6$ and (b) that on $ab$-plane with only Co-atoms. (c) First Brillouin zone for CoTa$_3$S$_6$ and (d) that on $k_z=0$ plane, where several high symmetry $\bm{k}$-points are shown by dots. }
    \label{fig:cryst_cota3s6}
\end{figure}

\begin{figure}
    \centering
    \includegraphics[width=1.0 \hsize]{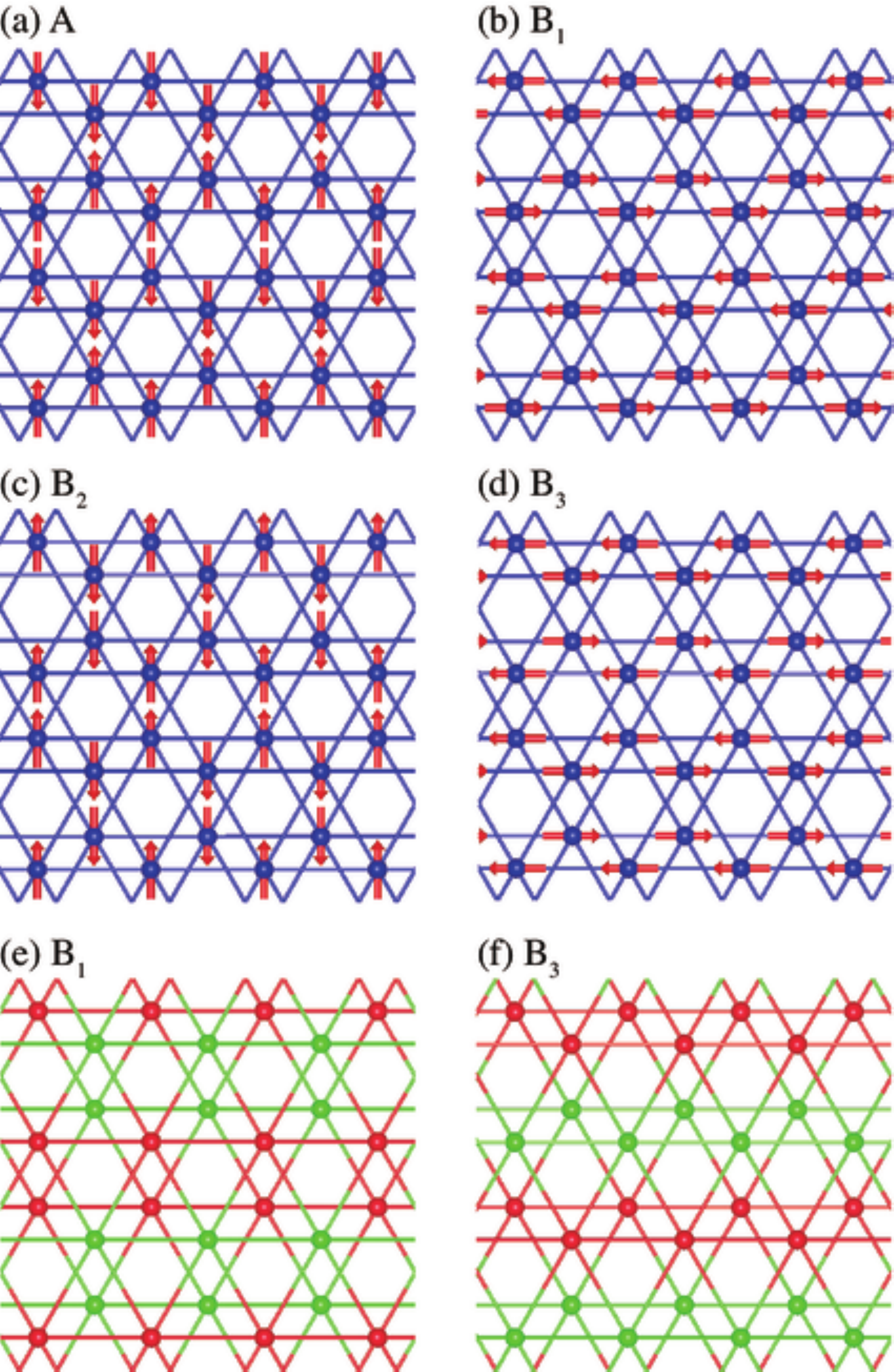}
    \caption{Magnetic structure bases with single-$\bm{k}$ $(0,\frac{1}{2},0)$ in 
     crystal structure of Co$M_3$S$_6$. 
    The arrow and red (green) sphere represents magnetic moment in $ab$-plane and along $+z$ ($-z$) axis, respectively.}. 
    \label{fig:single_q_M}
\end{figure}

\begin{figure*}
    \centering
    \includegraphics[width=1.0 \hsize]{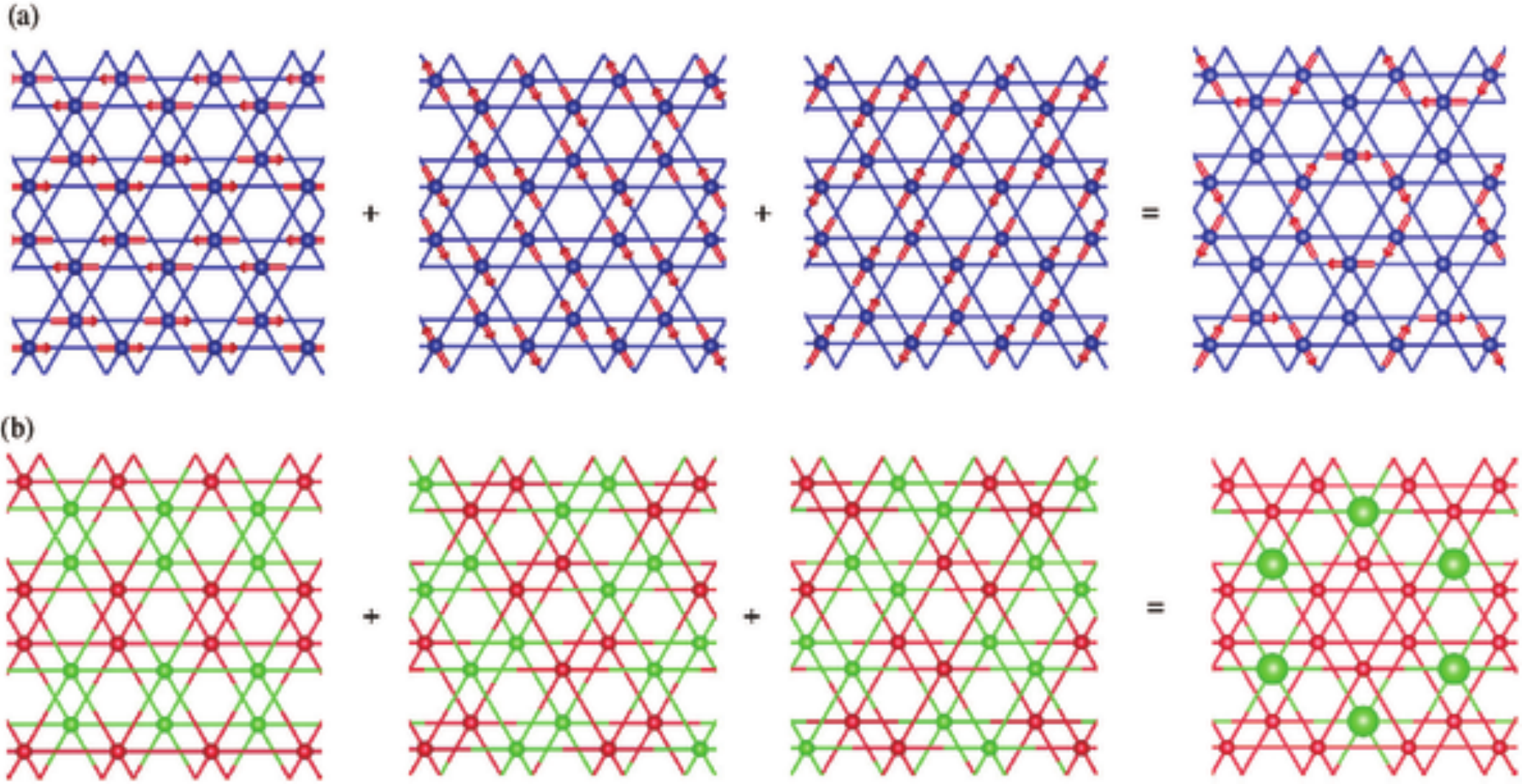}
    \caption{Triple-$\bm{k}$ magnetic structures induced from single-$\bm{k}$ magnetic bases with $B_1^-$ symmetry at M-points. 
    Panels~(a) and (b) correspond to those from magnetic bases shown in Figs.~\ref{fig:single_q_M}(b) and (e), respectively.}
    \label{fig:triple_q_M}
\end{figure*}
\begin{figure}
    \centering
    \includegraphics[width=1.0 \hsize]{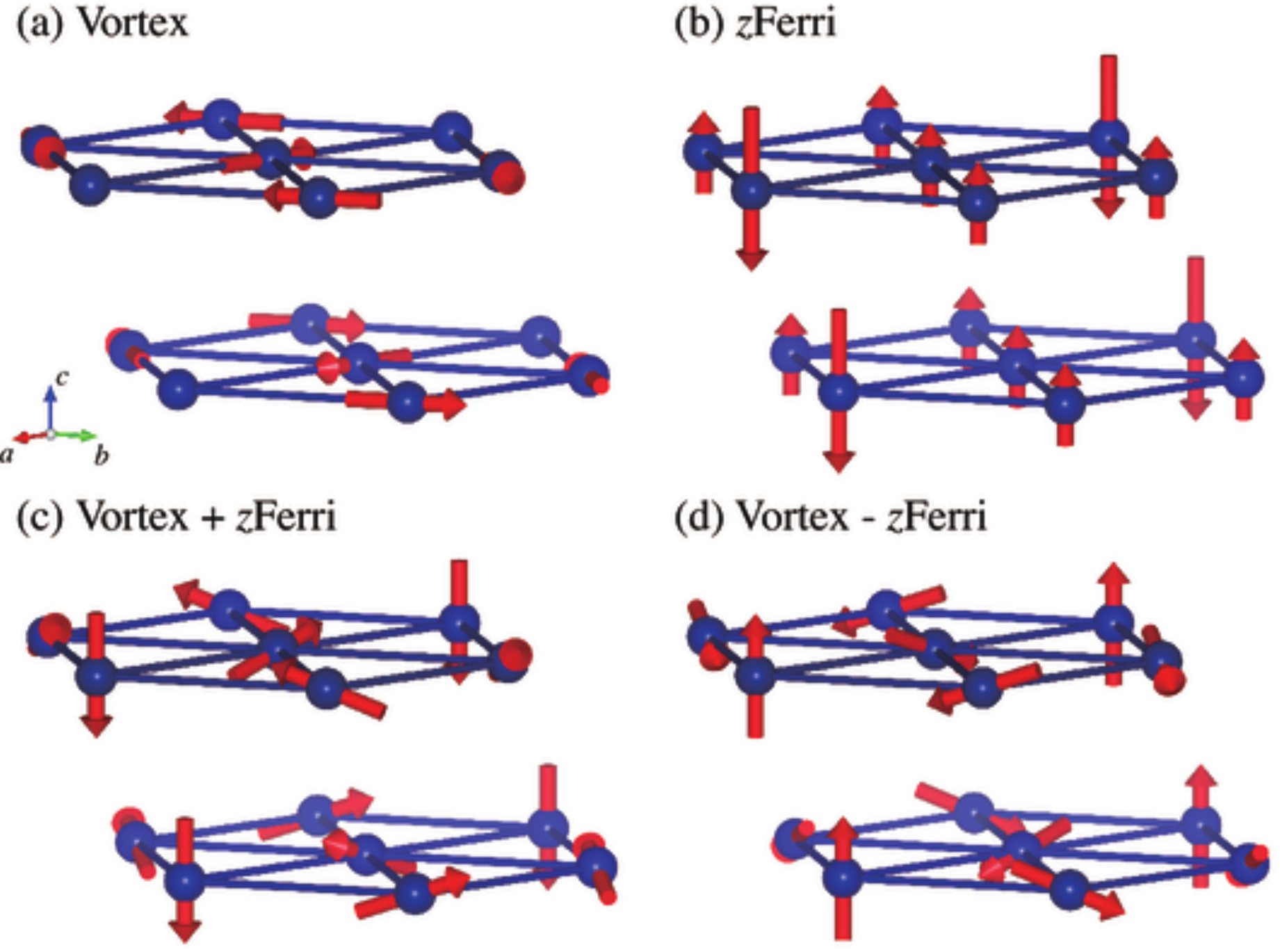}
    \caption{Examples of linear combinations of triple-$\bm{k}$ states shown in Fig.~\ref{fig:triple_q_M}.}
    \label{fig:triple_q_M_3D}
\end{figure}

An intercalated compound of transition metal  dichalcogenides, $TM_{3}$S$_{6}$ ($T$~=~$3d$-transition metal, $M$~=~Nb, Ta),  
 shows rich electronic and magnetic properties. For instance, a helimagnetic ordering with a long period is formed in Cr$M_3$S$_6$~\cite{Togawa_JPSJ.85.112001}, which is transformed into chiral soliton lattice under a magnetic field. Another compound, ferromagnetic VNb$_3$S$_6$ is proposed as a candidate of Weyl semimetals by first-principles calculations~\cite{Inoshita_PhysRevB.100.121112}.   
  CoNb$_3$S$_6$ in particular has attracted growing interests due to its large anomalous Hall conductivity in the magnetic ordered phase with a tiny ferromagnetic moment~\cite{Ghimire_Nat.Commun.9.3280,Tenasini_PhysRevResearch.2.023051,Mangelsen_PhysRevB.103.184408}.  
 This compound crystallizes in the space group $P6_322$ (No. 182, $D^{6}_6$) symmetry, where the crystal structure is shown  in Fig.~\ref{fig:cryst_cota3s6}. 
 Early neutron diffraction studies by Parkin \textit{et al.} indicate the magnetic structure with a single-$\bm{k}$   $\left(0,\frac{1}{2},0\right)$ 
 having multidomain for CoNb$_3$S$_6$ and 
 that with $\bm{k}=\left(\frac{1}{3},\frac{1}{3},0\right)$ for  CoTa$_3$S$_6$~\cite{Parkin_J.Phys.Solid.State.Phys.16.1765}.   
 Recent neutron diffraction patterns for  CoNb$_3$S$_6$, however, are also compatible with multi-$\bm{k}$ magnetic order with single domain~\cite{Tenasini_PhysRevResearch.2.023051}. We here discuss possible magnetic structures with ordering vector $\bm{k}$ at high symmetry points in BZ and provide a symmetry analysis of magnetic structures in connection with emergence of anomalous Hall effect.

 Let us consider magnetic structures with ordering vectors $\bm{k}$ at M-points. 
  There are three symmetry equivalent M-points in the first Brillouin zone, which are given by  
  $\bm{k}_{\mathrm{M}^{(1)}}=\left(0,\frac{1}{2},0\right)$, 
 $\bm{k}_{\mathrm{M}^{(2)}}=\left(-\frac{1}{2},0,0\right)$, 
 and $\bm{k}_{\mathrm{M}^{(3)}}=\left(\frac{1}{2},-\frac{1}{2},0\right)$ as shown in Figs.~\ref{fig:cryst_cota3s6}(c) and (d).  The $\bm{k}$-point group at M-points is $D_2$ and magnetic bases can be classified according to IRREPs under this point group (see Table~\ref{tab:character_M}).  
 Figure~\ref{fig:single_q_M} shows symmetry adapted magnetic structures with a single propagation vector $\bm{k} = \bm{k}_{\mathrm{M}^{(1)}}$. 
 All the magnetic structures in Fig.~\ref{fig:single_q_M} are collinear AFM. 
  Note that we can construct noncollinear magnetic structures as linear combinations of (b) and (e) and those of (d) and (f) without further symmetry lowering since the magnetic structures in Figs.~\ref{fig:single_q_M}(b) and (e) and those in (d) and (f) have same IRREPs.  Symmetry operation of time-reversal combined with primitive lattice translation holds in the magnetic states characterized by $\bm{k}=(0,\frac{1}{2},0)$ shown in Fig.~\ref{fig:single_q_M}. As a result, anomalous Hall response is prohibited in these magnetic states.

\begin{table}[t!]
\caption{
Character table for $\bm{k}$-group at $\mathrm{M}^{(i)}$-point.  
$C_{2x,1}$ and $C_{2y,1}$ represent two-fold rotation along $x$- and $y$-axes. 
$C_{2x(y),i}$ ($i=2,3$) is defined as $C_{2x(y),i}=C^{i-1}_{3z} C_{2x(y),1} C^{-i+1}_{3z}$. 
}
\label{tab:character_M}
\centering
\begin{tabular}{cccccc} \hline\hline
  & $E$ & $C_{2z}$ & $C_{2y,i} $  & $C_{2x,i}$ \\
 \hline
$A$   & $1$ & $1$  & $1$  & $1$ \\
\hline
$B_1$   & $1$ & $1$  & $-1$  & $-1$ \\
\hline
$B_2$   & $1$ & $-1$  & $1$  & $-1$ \\
\hline
$B_3$   & $1$ & $-1$  & $-1$  & $1$ \\
\hline\hline
\end{tabular}
\end{table}

  As mentioned in Sec.~II-C, triple-$\bm{k}$ magnetic structures are obtained from single-$\bm{k}$ magnetic bases. Possible high-symmetry triple-$\bm{k}$ magnetic structures related to anomalous Hall effect are shown in Fig.~\ref{fig:triple_q_M}.  
  The two magnetic structures, i.e., vortex-like arrangement of magnetic moments in $ab$-plane and ferrimagnetic structure along $c$-axis shown in Fig.~\ref{fig:triple_q_M}, can induce AHE since these magnetic point groups are $62'2'$~\cite{Seeman_PhysRevB.92.155138}. 
 The other high-symmetry triple-$\bm{k}$ magnetic states induced from single-$\bm{k}$ magnetic states having each of $A^-$, $B^-_2$, and $B^-_3$ symmetry in Table~\ref{tab:character_M}  prohibit AHE since these magnetic point groups are $622$, $6'2'2$, and $6'22'$, respectively~\cite{Seeman_PhysRevB.92.155138}. 
The Landau free energy expression gives more clear insight into emergence of AHE in magnetically ordered phase. 
 Secondary magnetic order parameters are coupled with the primary ones through the even order terms in the Landau free energy due to time-reversal symmetry in the paramagnetic state.
Magnetic multipoles with $\bm{k}=\bm{0}$ are coupled with magnetic order parameters  with $\bm{k}_{\mathrm{M^{(i)}}} $ through the fourth order terms in the Landau free energy since the sum of the wave 
 vectors at M-points is zero: $\bm{k}_{\mathrm{M}^{(1)}} + \bm{k}_{\mathrm{M}^{(2)}}+ \bm{k}_{\mathrm{M}^{(3)}} =\bm{0}$.    
 In the case that primary order parameters have $B^-_1$ symmetry with propagation vector at M-points, the fourth order term in the Landau free energy is given as follows: 
\begin{align}
F^{(4)}&
=
d_1 M_{B^{(1)}_1}   M_{B^{(2)}_1} M_{B^{(3)}_1} M_z \nonumber \\
 & +d_2 \left(M^2_{B^{(2)}_1} -M^2_{B^{(3)}_1} \right)M_{B^{(1)}_1}M_{A^{(1)}} \nonumber \\ 
 & +d_2 \left(M^2_{B^{(3)}_1} -M^2_{B^{(1)}_1} \right)M_{B^{(2)}_1}M_{A^{(2)}} \nonumber \\ 
 & +d_2 \left(M^2_{B^{(1)}_1} -M^2_{B^{(2)}_1} \right)M_{B^{(3)}_1}M_{A^{(3)}},    \label{eq:4th_Landau_CTS} 
\end{align}
where $M_{A^{(i)}}$ and $M_{B^{(i)}_{1}}$ represent the magnetic order parameters having $A^-$- and $B^-_{1}$-symmetry with ordering vector $\bm{k}=\bm{k}_{\mathrm{M}^{(i)}}$, respectively, and $M_z$ represents the magnetic order parameter having $A^{-}_2$-symmetry with ordering vector $\bm{k}=\bm{0}$, i.e., the order parameter with same symmetry for $z$-component of net magnetization. 
   The Eq.~(\ref{eq:4th_Landau_CTS}) clearly shows that uniform $M_z$ can be finite in the triple-$\bm{k}$ magnetic states induced from single-$\bm{k}$ magnetic states having $B^-_{1}$-symmetry, indicating the emergence of AHE $\sigma_{xy}\neq 0$ in the ordered phase though $M_z$ is not necessary to be finite. 
Note that magnetic toroidal dipole $T_z$ can be finite with $M_z$ since $T_z$ belongs to the same $A_2^-$-IRREP of $M_z$ in the present crystal structure having chiral point group $D_6$.  
The vortex-like arrangements of magnetic moments can be regarded as toroidal dipole moments~\cite{Spaldin_2008,Hayami_PhysRevB.90.024432}, 
 and this is consistent with the magnetic structure shown in Fig.~\ref{fig:triple_q_M}(a).  
 The existence of magnetic toroidal dipole moments allows the magneto-electric effect and non-reciprocal charge transport, and these macroscopic responses can be footprints of triple-$\bm{k}$ ordering in the present case. 

As mentioned in the previous subsection, the electric order parameters are coupled with magnetic ones through the odd order terms in Landau free energy. The third order terms are given as follows:  
\begin{align}
F^{(3)}
& 
=
c_1\left(2 M^2_{B^{(1)}_1}  
-  M^2_{B^{(2)}_1}-M^2_{B^{(3)}_1}\right) Q_{v}  \nonumber \\
  & + c_1\sqrt{3}  \left(  -  M^2_{B^{(2)}_1} + M^2_{B^{(3)}_1}\right) Q_{xy}  \nonumber\\
& +  c_2 M_{B^{(2)}_1}M_{B^{(3)}_1} Q_{A^{(1)}} \nonumber \\
& +  c_2 M_{B^{(3)}_1}M_{B^{(1)}_1} Q_{A^{(2)}} \nonumber \\
& +  c_2 M_{B^{(1)}_1}M_{B^{(2)}_1} Q_{A^{(3)}},\label{eq:3rd_Landau_CTS}
\end{align}
where $(Q_{xy},Q_{v})$ 
is the electric multipole having $E^+_2$-symmetry 
with propagation vector $\bm{k}=\bm{0}$ and $Q_{A^{(i)}}$ represents electric order parameter having $A^+$-symmetry with $\bm{k}=\bm{k}_{\mathrm{M}^{(i)}}$. 
 We can see that, from Eq.~(\ref{eq:3rd_Landau_CTS}), 
 single-$\bm{k}$ magnetic state $M_{B^{(1)}_1}\neq 0$ and $M_{B^{(2)}_1}=M_{B^{(3)}_1}=0$ break the crystallographic point group from hexagonal group $D_6$ to orthorhombic one $D_2$ and leads to the emergence of the secondary electric order parameter $Q_{v}$.
  The triple-$\bm{k}$ state, $M_{B^{(2)}_1}=M_{B^{(2)}_1}=M_{B^{(3)}_1}\neq 0$, preserves the  crystallographic point group $D_6$, and the electric multipoles with $\bm{k}=\bm{0}$ are not induced as a result.
 Note that the unit cell in the triple-$\bm{k}$ state is extended in the $ab$-plane from the nonmagnetic one though the triple-$\bm{k}$ state has the same crystallographic point group with that for the nonmagnetic state.   
 The resulting finite components of macroscopic response tensors associated with electric order parameters are thus unchanged from those in the paramagnetic phase. 
 Meanwhile, 
 the electric order parameters with $\bm{k}=\bm{k}_{\mathrm{M}^{(i)}}$ emerges 
 $ Q_{A^{(1)}}= Q_{A^{(2)}}= Q_{A^{(3)}}\neq 0$. 
  These order parameters $ Q_{A^{(i)}}$  corresponds to charge ordering with the propagation vector $\bm{k}_{\mathrm{M}^{(i)}}$ and  can be measured by the microscopic probes such as X-ray diffraction measurements.  
 Thus, the secondary order parameters are useful for gaining information about primary order ones also in this case. 
 

\section{Summary and outlook}

 We have proposed a generation method of symmetry-adapted magnetic structure basis set with finite propagation vector $\bm{k}$ based on multipole expansion. 
  In this method, the magnetic configurations on a virtual cluster of each propagation vector is first generated with use of multipole expansion and then these are mapped onto 
 crystallographic systems  
 with a phase factor to take into consideration of spatial modulation of magnetic structures specified by a wave vector $\bm{k}$. 
  The iterative implementation of this procedure with Gram-Schmidt orthogonalization provides complete orthonormal magnetic bases with a given $\bm{k}$ classified according to irreducible representations under $\bm{k}$-group  in arbitrary crystal structures. 
 We can also construct the multi-$\bm{k}$ magnetic structures as the induced representation 
  of
  the 
   single-$\bm{k}$ states.  The obtained magnetic structures can be good candidates for initial magnetic configurations for density functional theory (DFT) calculations~\cite{Huebsch_PhysRevX.11.011031} and thus high-throughput DFT calculations for magnetic materials combined with the present magnetic basis generation method is a possible future direction of our work. 

 The present method is applied  
 to 
 representative materials $\alpha$-Mn and Co$M_3$S$_6$  
 ($M$~=~Nb, Ta) 
  showing large anomalous Hall effect (AHE) with tiny net magnetization. 
 It is demonstrated that magnetic structures  compatible  
 with both AHE and small ferromagnetic moment can be constructed. 
 We also discuss the secondary order parameter,  
 physical responses,  
 and electronic properties 
 under magnetic orders in  $\alpha$-Mn and Co$M_3$S$_6$. 
 We show that
 various phenomena such as AHE, anomalous Nernst effect (ANE), and magneto-optical Kerr effect (MOKE) can emerge through the primary magnetic order parameter having $T_1^-$-symmetry with $\bm{k}=\bm{0}$  for high-pressure phase in $\alpha$-Mn, and the Edelstein effect driven by the odd parity electric order parameters are possible both in the ambient and high pressure phases since the symmetry of secondary electric order parameters are same  in both phases. 
 For Co$M_3$S$_6$,  the magnetic order parameter with $T_1^-$-symmetry with $\bm{k}=\bm{0}$ are induced as a secondary order parameter in triple-$\bm{k}$ state, 
  leading to emergence of AHE, ANE, MOKE, and non-reciprocal charge transport.  
 
 Our method is applicable to general magnets including noncollinear and noncoplanar antiferromagnets and skyrmion crystals~\cite{Nagaosa_Nat.Nanotech.8.899}. The present magnetic structure generation scheme thus would facilitate exploration of functional magnetic materials.

\begin{acknowledgments}
The authors thank K. Akiba, T. C. Kobayashi,  S. Araki, H. Fukazawa, N. Shioda, T. Ohama, and Y. Kohori for fruitful discussions on experimental data for $\alpha$-Mn.  
They are also grateful to S. Seki, H. Takagi, and S. Minami for collaborative research on Co$M_3$S$_6$. 
This research was supported by JSPS KAKENHI Grants Numbers 
JP15H05883 (J-Physics), JP18H04230 (Topological Materials Science), JP19H01842, JP19H05825 (Quantum Liquid Crystals), JP20H05262 (Hypermaterials), JP20K05299, JP20K21067, JP21H01031, JP21H01789, JP21H04437, and 
  JP21H04990, 
 by JST PRESTO Grants Numbers JPMJPR17N8 and JPMJPR20L7, and by JST-Mirai Program Grant Number JPMJMI20A1.
They acknowledge Center for Computational Materials Science, Institute for Materials Research, Tohoku University for the use of MASAMUNE-IMR. 
Figures of crystal and magnetic structures are created by using \textsc{vesta}~\cite{Momma:db5098}.
\end{acknowledgments}

\appendix

\section{Transformation properties of $\bm{\Psi}_{l\Gamma_{\bm{k}}\gamma}$}

In this appendix, we show  that $\bm{\Psi}^{(X)}_{l \Gamma_{\bm{k}}\gamma} $ in Eq.~(\ref{eq_magbase_cryst_k}) belongs to $\Gamma_{\bm{k}}$-IRREP. 
 Since magnetic bases on virtual cluster $\tilde{\Psi}_{l\Gamma_{\bm{k}}\gamma}$ have transformation properties of IRREP $\Gamma_{\bm{k}}$, we should show the following relation:   
  $ \{p_{i}h_{\zeta}|\bm{\tau}_{i}\}\bm{\Psi}^{(X)}_{l \Gamma_{\bm{k}}\gamma} =\sum_{\gamma'}  D^{\Gamma_{\bm{k}}}_{\gamma' \gamma} (\{p_{i}h_{\zeta}|\bm{\tau}_{i}\})  \bm{\Psi}^{(X)}_{l \Gamma_{\bm{k}}\gamma'}$, where $\{p_{i}h_{\zeta}|\bm{\tau}_{i}\}  \in \mathcal{G}_{\bm{k}}$.
 The basis $\bm{e}^{\mathrm{axial}}_{\bm{k} \alpha \mu}$ have the following transformation properties under a $\bm{k}$-group operation $\{p_{i}h_{\zeta}|\bm{\tau}_{i}\}$, 
 \begin{align}
&  \{p_{i}h_{\zeta}|\bm{\tau}_{i}\} \bm{e}^{\mathrm{axial}}_{\bm{k} \alpha \mu} =  \sum_{\nu}  D^{(\mathrm{axial})}_{\nu\mu} (p_{i}h_{\zeta})e^{-i \varphi^{\bm{k},\alpha}_{i_{\zeta}}} \bm{e}^{\mathrm{axial}}_{\bm{k} s_{i_{\zeta}}[\alpha] \nu}, \label{eq_eaxial_cryst}
   \end{align} 
   where $D^{(\mathrm{axial})}$ is the transformation matrix of axial vector.  We here note that  the following relation for $\bm{R}^{s_{I'}[\alpha]}_{I_{\bm{k}}} $ is satisfied, 
\begin{align}
\bm{R}^{s_{i_{\zeta}'}[\alpha]}_{i_{\zeta}}  & =  \{p_{i}h_{\zeta}|\bm{\tau}_{i}\}\bm{\eta}_{s_{i_{\zeta}'}[\alpha]} -    \bm{\eta}_{s_{i_{\zeta}}[s_{i_{\zeta}'}[\alpha]]} \nonumber \\
 & =  \{p_{i}h_{\zeta}|\bm{\tau}_{i}\}  \{p_{i'}h_{\zeta'}|\bm{\tau}_{i'}\} \bm{\eta}_{\alpha} - p_{i}h_{\zeta} \bm{R}^{\alpha}_{i_{\zeta}'} -    \bm{\eta}_{s_{i_{\zeta}''}[\alpha]} \nonumber  \\
 & =  \{p_{i''}h_{\zeta''}|\bm{\tau}_{i''}\} \bm{\eta}_{\alpha}  + \bm{T}- p_{i}h_{\zeta} \bm{R}^{\alpha}_{i_{\zeta}'} -    \bm{\eta}_{s_{i_{\zeta}''}[\alpha]}, \label{eq_RIk}
 \end{align}
where symmetry operation $p_{i''}h_{\zeta''}$ and a primitive translation  $\bm{T} $  
are given as follows: 
\begin{align}
&p_{i''}h_{\zeta''} = p_{i}h_{\zeta}p_{i'}h_{\zeta'},  \label{eq_I''}\\
& \bm{T} = p_{i}h_{\zeta} \bm{\tau}_{i'} + \bm{\tau}_{i} -\bm{\tau}_{i''}. \label{eq_T}  
 \end{align}
By taking account of the relations in Eqs.~(\ref{eq_RIk}), (\ref{eq_I''}), and (\ref{eq_T}), 
it is shown that 
$ \left\{p_{i}h_{\zeta}|\bm{\tau}_{i}\right\}\bm{\Psi}^{(X)}_{l \Gamma_{\bm{k}}\gamma}$ 
is written as follows: 
\begin{align}
&   \{p_{i}h_{\zeta}|\bm{\tau}_{i}\}  \bm{\Psi}^{(X)}_{l \Gamma_{\bm{k}}\gamma}  \nonumber \\
 &= \sum_{i'=1}^{N_{\mathrm{coset}}}  \sum_{\zeta'  =1}^{N_h} \sum_{\mu,\nu}    
 e^{-i\varphi^{\bm{k},1}_{i_{\zeta}''} - i \bm{k}\cdot  \bm{T} }  \nonumber \\
&\times  D^{(\mathrm{axial})}_{\mu\nu} (p_{i}h_{\zeta})u^{(X)}_{\bm{k}l \Gamma \gamma, i_{\zeta}' \nu}  \bm{e}^{\mathrm{axial}}_{\bm{k} s_{i_{\zeta}''} [1]\mu} . \label{eq_trans_psi1}
 \end{align}
Let us here derive the relation among $D^{\Gamma_{\bm{k}}}_{\gamma\gamma'}$, $D^{(\mathrm{axial})}_{\mu\nu}$, and $u^{(X)}_{\bm{k}l \Gamma \gamma, \alpha \nu}$ 
from the transformation properties of magnetic bases on virtual cluster $ \tilde{\bm{\Psi}}^{(X)}_{l \Gamma_{\bm{k}}\gamma} $. Similarly to Eq.~(\ref{eq_eaxial_cryst}), the magnetic bases 
 $\tilde{\bm{e}}^{\mathrm{axial}}_{\bm{k} i_{\zeta}' \mu}$
 are transformed under a  symmetry operation
 $\{p_{i}h_{\zeta}|\bm{\tau}_{i}\} $ 
as follows: 
 \begin{align}
&  \{p_{i}h_{\zeta}|\bm{\tau}_{i}\} \tilde{\bm{e}}^{\mathrm{axial}}_{\bm{k} i_{\zeta}' \mu} =  \sum_{\nu}  D^{(\mathrm{axial})}_{\nu\mu} (p_{i}h_{\zeta})e^{-i \tilde{\varphi}^{\bm{k},i_{\zeta}'}_{i_{\zeta}}} \tilde{\bm{e}}^{\mathrm{axial}}_{\bm{k} i_{\zeta}'' \nu}, 
   \end{align} 
where 
 the phase factor is given as 
 $\tilde{\varphi}^{\bm{k},i'_{\zeta}}_{i_{\zeta}}= \bm{k}\cdot \tilde{\bm{R} }^{i_{\zeta}'}_{i_{\zeta}}$ 
 with
 $ \tilde{\bm{R}}^{i_{\zeta}'}_{i_{\zeta}}  = \{p_ih_{\zeta}|\bm{\tau}_i\} \tilde{\bm{\eta}}_{i_{\zeta}'}- \tilde{\bm{\eta}}_{i_{\zeta}''}$. 
 We here use the definition of virtual cluster described in Sec.~II-B, that is, $\tilde{\bm{\eta}}_{i_{\zeta}'}= \{p_{i'}h_{\zeta'}|\bm{\tau}_{i'}\}\bm{r}_1 $ and the relation in Eq.~(\ref{eq_I''}). 
Therefore,  the magnetic basis $ \tilde{\bm{\Psi}}^{(X)}_{l \Gamma_{\bm{k}}\gamma} $ shows following transformation properties: 
\begin{align}
 &\left\{p_{i}h_{\zeta}|\bm{\tau}_{i}\right\} \tilde{\bm{\Psi}}^{(X)}_{l \Gamma_{\bm{k}}\gamma} \nonumber \\
& =  \sum_{i'=1}^{N_{\mathrm{coset}}}\sum_{\zeta'=1}^{N_h}\sum_{\mu\nu}  D^{(\mathrm{axial})}_{\mu\nu} (p_{i}h_{\zeta}) e^{-i\tilde{\varphi}_{i_{\zeta}}^{\bm{k},i_{\zeta}'}}
  u^{(X)}_{\bm{k}l \Gamma \gamma, i_{\zeta}' \nu}  \tilde{\bm{e}}^{\mathrm{axial}}_{\bm{k} i_{\zeta}''\mu}.  \label{eq_trans_psi_tilde}
\end{align}
By considering   the relation 
 $\left\{p_{i}h_{\zeta}|\bm{\tau}_{i}\right\} \tilde{\bm{\Psi}}^{(X)}_{l \Gamma_{\bm{k}}\gamma}= \sum_{\gamma'}D^{\Gamma_{\bm{k}}}_{\gamma'\gamma} (\{p_{i}h_{\zeta}|\bm{\tau}_{i}\}  ) \tilde{\bm{\Psi}}^{(X)}_{l \Gamma_{\bm{k}}\gamma'} $ 
 as well as  Eq.~(\ref{eq_trans_psi_tilde}), we can obtain the following relation: 
\begin{align}
& \sum_{\nu} D^{(\mathrm{axial})}_{\mu\nu} (p_{i}h_{\zeta}) u^{(X)}_{\bm{k}l \Gamma \gamma, \alpha \nu} e^{-i \tilde{\varphi}^{\bm{k},i_{\zeta}'}_{i_{\zeta}}} \nonumber \\
&= \sum_{\gamma'} D^{\Gamma_{\bm{k}}}_{\gamma'\gamma} (\{p_{i}h_{\zeta}|\bm{\tau}_{i}\}  )  u^{(X)}_{\bm{k}l \Gamma \gamma', i_{\zeta}'' \mu}. \label{eq_mtx_psi_tilde1}
\end{align}
 Putting Eq.~(\ref{eq_mtx_psi_tilde1}) into  Eq.~(\ref{eq_trans_psi1}), we obtain the following relation: 
\begin{align}
&  \left\{p_{i}h_{\zeta}|\bm{\tau}_{i}\right\}  \bm{\Psi}^{(X)}_{l \Gamma_{\bm{k}}\gamma}  \nonumber \\
 &=\sum_{\gamma'}  \sum_{i'=1}^{N_{\mathrm{coset}}}  \sum_{\zeta'=1}^{N_h}\sum_{\mu}  
 e^{-i\varphi^{\bm{k},1}_{i_{\zeta}''} - i \bm{k}\cdot  \bm{T}+i \tilde{\varphi}^{\bm{k},i_{\zeta}'}_{i_{\zeta}} }  \nonumber\\
 &\times D^{\Gamma_{\bm{k}}}_{\gamma' \gamma} (\{p_{i}h_{\zeta}|\bm{\tau}_{i}\} )  
u^{(X)}_{\bm{k}l \Gamma_{\bm{k}} \gamma', i_{\zeta}'' \mu} \bm{e}^{\mathrm{axial}}_{\bm{k} s_{i_{\zeta}''}[1]\mu}. 
 \end{align}
  Using the relations in Eqs.~(\ref{eq_I''}) and (\ref{eq_T}) and following the same manner of  calculation in Eq.~(\ref{eq_RIk}), 
 we can show that 
$\tilde{\bm{R}}^{i_{\zeta}'}_{i_{\zeta}}  =\bm{T}$ and 
$\tilde{\varphi}^{\bm{k},i_{\zeta}'}_{i_{\zeta}}=\bm{k}\cdot \bm{T}$. 
Consequently, the relation 
 $ \{p_{i}h_{\zeta}|\bm{\tau}_{i}\}\bm{\Psi}^{(X)}_{l \Gamma_{\bm{k}}\gamma} =\sum_{\gamma'}  D^{\Gamma_{\bm{k}}}_{\gamma' \gamma} (\{p_{i}h_{\zeta}|\bm{\tau}_{i}\})  \bm{\Psi}^{(X)}_{l \Gamma_{\bm{k}}\gamma'}$ 
holds and thus, the magnetic basis $\bm{\Psi}^{(X)}_{l \Gamma_{\bm{k}}\gamma}$ belongs to $\Gamma_{\bm{k}}$-IRREP under $\bm{k}$-group $\mathcal{G}_{\bm{k}}$.

\bibliography{ref}

\begin{thebibliography}{77}
\bibitem [{pre()}]{previous_vc}%
  \BibitemOpen
  \href@noop {} {}\bibinfo {note} {In other words, we decompose a virtual
  cluster obtained by the method in Ref.~\cite{Suzuki_PhysRevB.99.174407} into
  $N_{\mathrm{coset}}$ clusters by shifting those atomic positions
  $p_{i}h_j\bm{r}_1$ by $\bm{\tau}_i$.}\BibitemShut {Stop}%
\end{thebibliography}

\end{document}